\documentclass[12pt]{iopart}

\newcommand{\eq}[1]{\begin{equation}#1\end{equation}}

\newcommand{\ee}{\mathrm{e}}
\usepackage{iopams}
\usepackage{graphics}
\usepackage{graphicx}

\begin{document}

\title{Reduced density matrices and entanglement entropy in free lattice
models}

\author{ Ingo Peschel$^1$ and Viktor Eisler$^2$}
\address{$^1$ Fachbereich Physik, Freie Universit\"at Berlin, Arnimallee
 14, D-14195 Berlin, Germany\\
$^2$ Niels Bohr Institute, University of Copenhagen, Blegdamsvej 17,
DK-2100 Copenhagen \O, Denmark}

%\maketitle

\begin{abstract}
We review the properties of reduced density matrices for free fermionic or bosonic 
many-particle systems in their ground state. Their basic feature is that they have a 
thermal form and thus lead to a quasi-thermodynamic problem with a certain free-particle
Hamiltonian. We discuss the derivation of this result, the character of the Hamiltonian
and its eigenstates, the single-particle spectra and the full spectra, the resulting
entanglement and in particular the entanglement entropy. This is done for various one- 
and two-dimensional situations, including also the evolution after global or local 
quenches. 

%Key words :

\end{abstract}

\section{Introduction}

Reduced density matrices contain the information on some part of a quantum
system and are a basic tool in many-body physics. The ones commonly employed
describe the properties of one or two selected \emph{particles} in a many-particle
system and allow to calculate important physical quantities like the total
energy or the density correlations. These reduced density matrices (RDM's) were 
first introduced by Dirac \cite{Dirac30} and studied already in the 1930's, 
see e.g. \cite{Fock31,Watanabe39}. In the usual terminology, they are just the 
static one- and two-particle correlation functions.

The RDM's we want to discuss here are of a different type and refer to a different
question. They arise if one divides a system in \emph{space}, or, more generally, 
in Hilbert space, and asks how the two parts are coupled in the given wave function. 
This corresponds to the analysis by Schr\"odinger in 1935 \cite{Schroedinger35} 
when he introduced the concept of entanglement. The general form of this coupling is 
given by the Schmidt decomposition which displays all entanglement features 
in a simple and transparent way. To obtain it in a specific case, one needs the RDM's 
for the two \emph{regions} in question.

The present interest in this problem, although it had also been a theme in quantum optics,
arose in the beginning of the nineties in two seemingly disconnected areas, in the
theory of black holes \cite{Bombelli86,Srednicki93,Holzhey94} and in the numerical 
investigation of quantum chains \cite{White92,White93}. In both cases, the motivation
came from the wish to consider some subsystem which is in contact with its environment.
For the quantum chains, this lead to the density-matrix renormalization group (DMRG) 
which can treat large systems with spectacular accuracy and revolutionized the field 
\cite{Buch99,Schollwoeck05}. A third input then came from the area of 
quantum information, where the structure of quantum states also plays a central role.
This resulted in particular in a renewed and extensive study of the entanglement 
entropy \cite{Vidal03,Latorre04,Amico08} which is a simple and convenient measure 
of the entanglement and follows directly from the RDM eigenvalues.

The purpose of this article is to give a coherent account of the reduced density matrices
just described for a class of models where they can be obtained in closed form. 
These are free fermions including the related spin chains and free bosons in the form 
of coupled oscillators. They will be considered either in their ground state or in certain 
other pure quantum states. In this case, the RDM's are found to have a Boltzmann-like
form with a certain free-particle operator in the exponent. The problem is thereby
reduced to the study of this associated Hamiltonian and its characteristic features. 
The main property of interest is the eigenvalue spectrum since it determines the
spectrum of the RDM itself and thus the entanglement properties, in particular the
entanglement entropy. Both the spectra and the entropies will be presented for a variety 
of different situations. The problem on a lattice is very clear-cut. The partitioning is 
done by selecting two sets of discrete sites and there are no divergencies for finite
sizes. On the other hand, there is only a small number of analytical results and one
has to invoke numerics frequently. Lattice systems are also required for the DMRG, 
and the initial motivation for the studies was to understand the performance of this 
intriguing numerical method by looking at solvable models.

In section 2 we will provide some background on entanglement, the Schmidt decomposition 
and the RDM's. Then, in section 3, we give the general form of the reduced density
matrices for free fermions or bosons and discuss the methods for obtaining them. 
For quantum chains, this also contains relations to two-dimensional classical models. 
In section 4 we show the eigenvalue spectra for various one- and two-dimensional systems 
and discuss their typical appearance, their scaling behaviour and the change with the
dimension. The characteristics of the single-particle eigenfunctions, the nature of the
effective Hamiltonian and some further aspects are the topics of section 5.
In section 6 we turn to the entanglement entropy and summarize the important results 
with emphasis on their relation to the spectra. Finally, in section 7, we review 
the temporal behaviour of the entanglement after different types of quenches. The 
material is drawn preferentially from our own studies and some of it also appeared  
in a recent book \cite{Greifswald08}. However, the scope is different here and a 
considerable number of figures were prepared exclusively for this review.

%
%--------------------------------------------------------------------------------
%
\section{Background}

In this section, we summarize the basic features of entangled states and reduced
density matrices in order to create the frame for the results to be presented later.
For more details, see e.g. the short review \cite{Ekert/Knight95}.

\subsection{Schmidt decomposition}

Consider a quantum system which is divided into two distinct parts 1 and 2. 
Then a state $|\Psi\rangle$ of the total system can be written 
\begin{equation}
 |\Psi\rangle = \sum_{m,n} A_{m,n} |\Psi^1_m\rangle  |\Psi^2_n\rangle
\label{state}
\end{equation}
where $|\Psi^1_m\rangle$ and $|\Psi^2_n\rangle$ are orthonormal 
basis functions in the
two Hilbert spaces. But a rectangular matrix $\bf{A}$ can always be written 
in the form $\bf{UDV'}$ where $\bf{U}$ is unitary, $\bf{D}$ is diagonal and 
the rows of $\bf{V}$ are orthonormal. This is called the singular-value 
decomposition and similar to the principal-axis transformation of a symmetric 
square matrix \cite{Horn/Johnson91}. Using this in (\ref{state}) and forming
new bases by combining the $|\Psi^1_m\rangle$ with $\bf{U}$
and the $|\Psi^2_n\rangle$ with $\bf{V'}$, one obtains the 
Schmidt decomposition \cite{Schmidt07} 
\begin{equation}
 |\Psi\rangle = \sum_{n}\lambda_{n}\,|\Phi^1_n\rangle  |\Phi^2_n\rangle
\label{schmidt}
\end{equation}
which gives the total wave function as a single sum of products of orthonormal 
functions. Here the number of terms is limited by the smaller of the two Hilbert 
spaces and the weight factors $\lambda_{n}$ are the elements of the diagonal matrix 
$\bf{D}$. If $|\Psi\rangle$ is normalized, their absolute magnitudes squared sum to one.
The entanglement properties are encoded in the set of $\lambda_{n}$. Only if all
except one are zero, the sum reduces to a single term and $|\Psi\rangle$ is a product
state, i.e. non-entangled. In all other cases a certain entanglement is present
and if all $\lambda_{n}$ are equal in size, one would call the state maximally 
entangled. Of course, this refers to a particular bipartition and one should investigate 
different partitions to obtain a complete picture.

\subsection{Reduced density matrices}

The entanglement structure just discussed can also be found from the density
matrices associated with the state $|\Psi\rangle$. This is, in fact, the standard
way to obtain it. Starting from the total density matrix
\begin{equation}
 \rho =  |\Psi\rangle \langle\Psi|
\label{rhotot}
\end{equation}
one can, for a chosen division, take the trace over the degrees of freedom in one
part of the system. This gives the reduced density matrix for the other part, i.e.
\begin{equation}
\rho_{1} = \mathrm{tr}_{2}(\rho) \;\;,\;\; \rho_{2} = \mathrm{tr}_{1}(\rho)
\label{rho12}
\end{equation}
These hermitian operators can be used to calculate arbitrary expectation values in
the subsystems. Moreover, it follows from (\ref{schmidt}) that their 
diagonal forms are
\begin{equation}
 \rho_{\alpha} = \sum_{n}|\lambda_{n}|^2\; |\Phi^{\alpha}_{n}\rangle 
 \langle\Phi^{\alpha}_{n}|\;\;\;,\;\;
 \alpha=1,2
\label{rhodiag}
\end{equation}
This means that
\begin{itemize}
\item $\rho_{1}$ and $\rho_{2}$ have the same non-zero eigenvalues
\item these eigenvalues are given by $w_{n}=|\lambda_{n}|^2$
\end{itemize}
Therefore the eigenvalue spectrum of the $\rho_{\alpha}$ gives directly the weights
in the Schmidt decomposition and a glance at this spectrum shows the basic entanglement
features of the state, for the chosen bipartition. For this reason, it has also been
termed ``entanglement spectrum'' recently \cite{Li/Haldane08}. One also sees that the 
$|\Phi^{\alpha}_{n}\rangle$ appearing in (\ref{schmidt}) are the eigenfunctions of the
$\rho_{\alpha}$. For the single-particle RDM's mentioned in the introduction, these
eigenfunctions are known as ``natural orbitals'' in quantum chemistry \cite{Loewdin55}.
\par
In the DMRG algorithm, these properties are used to truncate the Hilbert space by 
calculating the $\rho_{\alpha}$, selecting the $m$ states $|\Phi^{\alpha}_{n}\rangle$ 
with largest weights $w_n$ and deleting the rest. This procedure is expected to work 
well if the total weight of the discarded states is sufficiently small. Therefore the 
form of the density-matrix spectra is decisive for the success of the method.
\par
It is interesting that Schmidt himself already worked with the RDM's. Studying
coupled linear integral equations, he derived a spectral representation of the form  
(\ref{schmidt}) for an unsymmetric kernel $K$ in terms of the eigenfunctions of the
two symmetric operators $KK'$ and $K'K$. His paper (which is based on his doctoral
thesis with Hilbert) also contains the recipe for the best approximation as it is 
used in the DMRG.
%
%------------------------------------------------------------------
%
\subsection{Entanglement entropy}

Whereas the full RDM spectra give the clearest impression of the entanglement
in a bipartite system, it is also desirable to have a simple measure which
condenses this information into one number. This can be achieved by
generalizing the usual (von Neumann) entropy definition to reduced density 
matrices. The entanglement entropy therefore reads:
\begin{equation}
S_1 = - \mathrm{tr}(\rho_1 \ln \rho_1) = - \sum_n w_n \ln w_n,
\label{ent_def}
\end{equation}
where the trace has been rewritten as a sum using the eigenvalues $w_n$.
The most important properties are as follows.
\begin{itemize}
\item The entropy is determined purely by the spectrum of $\rho_1$,
which is known to be identical to the spectrum of $\rho_2$.
Therefore $S_1=S_2$ holds for arbitrary bipartitions and one can simply
write $S$ and talk of \emph{the} entanglement entropy.

\item The entropy vanishes for product states,
and has a maximal value of $S=\ln M$ if one has $M$ non-zero eigenvalues which
are all equal, $w_n= 1/M$ for $n=1,2,\dots,M$. Using this, one can write 
in general $S=\ln M_{\mathrm{eff}}$, thereby defining an effective number of states 
coupled in parts $1$ and $2$. This gives a simple interpretation to $S$.
\end{itemize}
\par
Although there are other entanglement measures \cite{Plenio/Virmani07}, the entropy is the 
standard one for bipartitions and will be discussed in detail later. It is important 
to keep in mind that it measures a \emph{mutual connection} and will, in general,
not be proportional to the size of a subsystem.

\par

%
%--------------------------------------------------------------------------------
%
\section{RDM's for free lattice models}

\subsection{Systems}

In the following we consider models with a Hamiltonian which is quadratic
in either fermion or boson operators and thus can be diagonalized by a
Bogoliubov transformation. In principle, these can be quite general, but 
we will concentrate on the following physically important systems 

\begin{itemize}
\item Fermionic hopping models with conserved particle number and Hamiltonian
\begin{equation} 
H=- \frac {1}{2} \sum_{<m,n>}  t_{m,n} c_m^{\dagger} c_n 
\label{hopping}
\end{equation}
where the symbol $< >$ denotes nearest neighbours. Apart from homogeneous
systems we will consider dimerized chains, where $t_{n,n+1}$ alternates
between $1 \pm \delta$, and the case of single defects.
\item Coupled oscillators with eigenfrequency $\omega_0$ and Hamiltonian
\begin{equation}
H= \sum_n \left[- \frac {1}{2} \frac {\partial^2}{\partial x^2_n}+
\frac {1}{2} \omega^2_0 x^2_n  \right] + \frac {1}{4} \sum_{<m,n>} k_{m,n} (x_m- x_n)^2
\label{oscillators}
\end{equation}
These are systems with an optical spectrum and bosonic pair creation and 
annihilation.
\item Spin one-half chains which are equivalent to free fermions via the 
Jordan-Wigner transformation. The most general one is the XY chain with a Z field,
described by
\begin{equation}
H= - \sum_n \left[ \frac {1+\gamma}{2} \sigma^x_n \sigma^x_{n+1}
+ \frac {1-\gamma}{2} \sigma^y_n \sigma^y_{n+1} \right]  - h \sum_n \sigma^z_n
\label{XY}
\end{equation}
where the $\sigma^{\alpha}_n$ are Pauli matrices at site $n$. For $\gamma=0$ this
reduces to the XX model, corresponds to (\ref{hopping}) with nearest-neighbour
hopping and can also model hard-core bosons. For $\gamma \ne 0$, it contains pair 
creation and annihilation terms. For $\gamma=1$ it becomes the Ising model in a 
transverse field (TI model) which we write, in a slightly different notation 
\begin{equation}
H= - \sum_n \sigma^z_n - \lambda \sum_n \sigma^x_n \sigma^x_{n+1},
\label{TI}
\end{equation}
\end{itemize}
\par
The solubility of the models in itself does not yet mean that the RDM's are easily
accessible. For example, they have been considered in the critical XXZ spin chain, 
but the formulae are very complicated, see \cite{Sato06,Sato07}. The free lattice models, 
however, have eigenstates with special properties which permit to make a simple 
general statement.

\subsection{General result}

For these free-particle models, the reduced density matrices for the ground state
can be written
\begin{equation}
% \fbox{\parbox{6.5cm}{\[
\rho_{\alpha} = \frac{1}{Z}\; e^{-\mathcal{H}_{\alpha}} \; , \quad
{\mathcal{H}_{\alpha}} = \sum_{l=1}^L \varepsilon_l f_l^{\dagger} f_l 
%\] }}
\label{rhogen}
\end{equation}
Here $L$ is the number of sites in subsystem $\alpha$ and 
the operators $f_l^{\dagger}$ ,$f_l$ are fermionic or bosonic creation and 
annihilation operators for single-particle states with eigenvalues $\varepsilon_l$.
The $f$'s are related to the original operators in the subsystem by a canonical 
transformation. Thus $\rho_{\alpha}$ has the form of a thermal density matrix with an 
effective Hamiltonian $\mathcal{H}_{\alpha}$ which is of the same free-particle type as $H$. 
In (\ref{rhogen}) it is already given in diagonal form. The constant $Z$, written in 
analogy to thermodynamics, ensures the correct normalization 
$\mathrm{tr}(\rho_{\alpha})=1$.
\par
This form of $\rho_{\alpha}$ is rather suggestive since one has a similar situation as for
a system in contact with a thermal bath. However, no assumption about the relative 
sizes of the two coupled systems enters here. More importantly, the operator 
$\mathcal{H}_{\alpha}$ is \emph{not}
the Hamiltonian $H$ restricted to the subsystem $\alpha$. Therefore (\ref{rhogen}) is
not a true Boltzmann formula. Nevertheless, the problem has been reduced to the study
of a certain Hamiltonian and its thermodynamic properties. The features of 
$\mathcal{H}_{\alpha}$ will be the topic of the next chapters. Generally, one can say
that it corresponds to an inhomogeneous system even if the subsystem it describes
is homogeneous. This will be seen in more detail in section 5.2. Here we first discuss how 
one arrives at (\ref{rhogen}). These considerations will also show that validity of 
(\ref{rhogen}) goes even beyond the ground state.  
\subsection{Methods}

Basically, there are three methods to obtain the reduced density matrices.

\bigskip\par
(I) Integration over part of the variables according to the definition (\ref{rho12}). 
This can be done e.g. for $N$ coupled harmonic oscillators \cite{Peschel/Chung99,
Chung/Peschel00}. In this case the ground state is a Gaussian in the normal coordinates, 
provided no normal frequency vanishes. In terms of the original coordinates $x_n$ of the 
oscillators, it has the form
\begin{equation}
\Psi(x_1,x_2,\dots,x_N)=C\; \exp(-\frac {1}{2}\sum_{m,n}^{N} A_{m,n}\; x_m\; x_n)
\label{gauss}
\end{equation}
Here $C$ is a normalization constant and the matrix $\mathcal A$ is the square root 
$\mathcal V^{1/2}$ of the dynamical matrix associated with the potential energy.
By forming $\rho$ and integrating out e.g. the variables $x_{L+1},\dots,x_N$ one obtains
$\rho_1(x_1,x_2,\dots,x_L\,|\,x'_1,x'_2,\dots,x'_L)$ which is again a Gaussian.
With proper linear combinations $y_l$ of the coordinates, it contains only 
squares $y_l^2, {y'_l}^2$ and differences $(y_l-y'_l)^2$. Early treatments worked with this
integral operator \cite{Bombelli86,Srednicki93}. However, one can convert the differences
into second derivatives and thereby obtain the differential operator
\begin{equation}
\rho_1= K \prod_{l=1}^L \,\exp(-\frac {1}{4}\omega_l^2 y_l^2)
\,\exp(\;\frac {1}{2} \;\frac {\partial^2}{\partial y_l^2})
 \,\exp(-\frac {1}{4}\omega_l^2 y_l^2)
\label{derivative}
\end{equation}
where the exponents become quadratic expressions in terms of boson operators. A 
diagonalization then gives the single exponential (\ref{rhogen}) with $\mathcal{H}_{1}$
describing a collection of $L$ new harmonic oscillators. Their eigenfrequencies 
$\varepsilon_l$ follow from $\mathcal A$ by dividing it into the submatrices
$a^{11},a^{12},a^{21}$, and $a^{22}$, according to whether the sites are 
in part 1 or in part 2. Then the $L \times L$ matrix 
$a^{11}(a^{11}-a^{12}(a^{22})^{-1}a^{21})^{-1}$
has the eigenvalues $\coth^2(\varepsilon_l/2)$.
\par
If $L=1$, there is just one such
oscillator with a frequency $\varepsilon$ which differs from $\omega_0$.  
Its eigenstates have a different spatial extent and may therefore be called 
``squeezed''. For $N=2$ the resulting
Schmidt decomposition of $\Psi(x_1,x_2)$ in terms of these states can easily be 
written down and is well known, see e.g. \cite{Han/Kim/Noz99,Braunstein/Loock05}.
\par
The method can also be used for systems of non-interacting fermions.  
In this case one first has to write the ground state in exponential form and then use 
Grassmann variables for the integration \cite{Chung/Peschel01,Cheong/Henley04/1}.
\bigskip\par
(II) Via correlation functions \cite{Peschel03,Vidal03,Latorre/Riera09}.
The simplest case is a system of free electrons hopping on $N$ lattice sites in a state 
described by a Slater determinant. In such a state, all many-particle correlation 
functions factorize into products of one-particle functions. For example, 
\begin{equation}
\langle c_m^{\dagger}c_n^{\dagger}c_kc_l\rangle=
\langle c_m^{\dagger}c_l\rangle\langle c_n^{\dagger}c_k\rangle-
\langle c_m^{\dagger}c_k\rangle\langle c_n^{\dagger}c_l\rangle
\label{factor}
\end{equation}
If all sites are in the same subsystem, a calculation using the reduced density matrix
must give the same result. This is guaranteed by Wick's theorem if $\rho_{\alpha}$
is the exponential of a free-fermion operator 
\begin{equation}
 \rho_{\alpha} = K \exp{(-\sum_{i,j=1}^{L} h_{i,j} c_i^{\dagger} c_j )}
\label{expo2}
\end{equation}
where $i$ and $j$ are sites in the subsystem. With the form of $\rho_{\alpha}$ fixed,
the hopping matrix $h_{i,j}$ is then determined such that it gives the correct
one-particle correlation functions $C_{i,j}=\langle  c_i^{\dagger} c_j\rangle$. 
The two matrices are diagonalized by the same transformation and one finds
(see also \cite{Cheong/Henley04/1})
\begin{equation}
\bf{h} = \ln{\,[(\bf{1}-\bf{C})/\bf{C}\,]}
\label{corr}
\end{equation}
The same formula also relates the eigenvalues $\varepsilon_l$ and $\zeta_l$ of 
${\bf{h}}$ and ${\bf{C}}$.
Expressed differently, the $\varepsilon_l$ follow from the equation
\begin{equation}
({\bf{1}}-2{\bf{C}})\,\phi_l = \tanh(\frac{\varepsilon_l}{2})\, \phi_l.
\label{purehopping}
\end{equation}
\par
If there is pair creation and annihilation, one has to include the 'anomalous' 
correlation functions $F_{i,j}=\langle c_i^{\dagger} c_j^{\dagger}\rangle$ and 
$F^*_{i,j}=\langle c_j c_i\rangle$. To reproduce them, the operator $\mathcal{H_{\alpha}}$ 
then must also contain pair terms. Diagonalizing it in the usual way \cite{LSM61},
one finds that the single-particle eigenvalues follow from two coupled equations,
which can be combined into a single one. For real $\bf{F}$ this reads
\begin{equation}
(2{\bf{C}}-{\bf{1}}-2{\bf{F}})(2{\bf{C}}-{\bf{1}}+2{\bf{F}})\,\phi_l =
\tanh^2(\frac{\varepsilon_l}{2})\, \phi_l.
\label{pair}
\end{equation}
and reduces to the previous result  (\ref{purehopping}) if $\bf{F}$ vanishes.
Alternatively, one can work with Majorana operators \cite{Vidal03,Latorre04}
$a_{2n-1}=(c_n+c_n^{\dagger})$ and $a_{2n}=i(c_n-c_n^{\dagger})$
and form the $2N \times 2N$ correlation matrix $M_{m,n}=\langle  a_m a_n\rangle$.
Restricted to the subsystem, it contains the same elements as the two matrices in 
(\ref{pair}) but arranged differently. Writing $M_{m,n}=\delta_{m,n}+i\Gamma_{m,n}$, 
the matrix $\Gamma$ of the subsystem has the eigenvalues $\pm i\tanh(\varepsilon_l
/2)$. 
\par
This method is very general. It works in any dimension, for arbitrary quadratic
Hamiltonians, for all states which are Slater determinants, and even at finite
temperature. Thus it has been used in a large number of situations ranging from 
homogeneous chains to defect problems, random systems, higher dimensions and the 
time evolution after a quench.  
\par
Factorization properties as in  (\ref{factor}) are well-known for Gaussians,
and therefore the approach is equally applicable to coupled oscillators in the ground
state (\ref{gauss}). Thus $\rho_{\alpha}$ must be the exponential of a bosonic
operator (as found in (I)) and  $\mathcal{H}_{\alpha}$ is again determined such that it 
reproduces the correlation functions, in this case those of positions and momenta,
$X_{i,j}=\langle x_i x_j\rangle$ and  $P_{i,j}=\langle p_i p_j\rangle$. In analogy to 
(\ref{pair}) the single-particle eigenvalues 
then follow from \cite{Audenaert02,Peschel03,Cramer06}
\begin{equation}
2 {\bf{P}}\;2 {\bf{X}}\;\phi_l=\coth^2(\frac{\varepsilon_l}{2})\;\phi_l.
\label{boson}
\end{equation}
Since for the total system $2\mathcal P =\mathcal V^{1/2} =\mathcal A $ and 
$2\mathcal X=\mathcal V^{-1/2}$, the matrix on the left side of (\ref{boson}) is 
seen to be the restriction of $\mathcal A$ to the subsystem multiplied by the
restriction of its inverse. This is exactly the expression given in (I).
As in the fermionic case, one can also combine coordinates and momenta,
which are analogous to the Majorana variables, and consider the corresponding
$2L \times 2L$ correlation matrix, usually called covariance matrix. Its reduction
to diagonal form is a well-known problem in mathematics \cite{Williamson36}
and the resulting $\coth(\varepsilon_l/2)$ are also referred to as symplectic 
eigenvalues \cite{Botero/Reznik04,Adesso/Illuminati07}. 
\par
The method was used for example in \cite{Casini/Huerta05,BCS06,Cramer06} and
again works also at finite temperature. 
\bigskip\par
(III) Via classical statistical models \cite{Nishino/Okunishi97,PKL99}.
In one dimension one can exploit the relations between quantum chains and 
two-dimensional classical models. The starting point is a discrete version of
a path-integral representation.

\begin{figure}[htb]
\centering
\includegraphics[scale=.7]{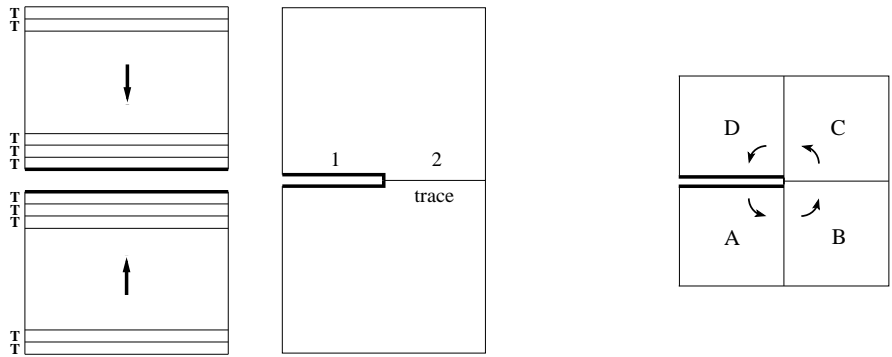}
\caption{Left: Density matrices for a quantum chain as two-dimensional partition
functions. Far left: Expression for $\rho$. Half left: Expression for $\rho_1$.
The matrices are defined by the variables along the thick lines.
Right: Two-dimensional system built from four quadrants with
corresponding corner transfer matrices $A,B,C,D$. The arrows indicate 
the direction of transfer. After Ref. \cite{Greifswald08}.}
\label{fig:pathint}
\end{figure}

Consider a quantum chain of finite length and imagine that one can obtain its
state $|\Psi\rangle$ from an initial state $|\Psi_s\rangle$ by applying a proper 
operator $T$ many times. If $T$ is the row-to-row transfer matrix of a classical model, 
one has thereby related $|\Psi\rangle$ to the partition function of a two-dimensional
semi-infinite strip of that system. The total density matrix $|\Psi\rangle \langle\Psi|$ 
is then given by two such strips. This is sketched on the far left of Fig.\ref{fig:pathint}.
The reduced density matrix, e.g. for the left part of the chain, follows by identifying
the variables along the right part of the horizontal edges and summing them, which means
tying the two half-strips together. In this way, $\rho_{\alpha}$ is expressed as the 
partition function of a full strip with a perpendicular cut, as shown half left in the 
figure.
\par
%
%%%%%%%%%%%%%%%%%%%%%%%%%%%%%%%%%%%%%%%%%%%%%%%%%%%%%%%%%%%%%%%%%%%%%%%%%%%%
%%%%%%%%%%%%%%%%%%%%%%%%%%%%%%%%%%%%%%%%%%%%%%%%%%%%%%%%%%%%%%%%%%%%%%%%%%%%%%%%

This procedure works for the ground state of a number of integrable quantum chains. 
For example, the TI chain can in this way be related to a two-dimensional Ising model 
on a square lattice which is rotated by $45^\circ$ with respect to the horizontal 
\cite{PKL99}. In the same way, a chain of coupled oscillators is connected with a 
two-dimensional Gaussian model \cite{Peschel/Chung99} and an XY chain with an
Ising model on a triangular lattice \cite{XYPeschel04}. Analogous correspondences link
XXZ, XYZ and higher-spin chains to vertex models \cite{PKL99,Ercolessi09,Weston06}.
To use these relations, however, one needs a way to actually calculate the resulting
partition function.
This is possible with the help of the corner transfer matrices (CTM's) introduced by Baxter 
\cite{Baxter82}. These are partition functions of whole quadrants as shown on the right 
of Fig.\ref{fig:pathint}, or of sextants, if one is dealing with a triangular lattice.
By multiplying these transfer matrices one can then obtain the reduced 
density matrix for a half-chain as
\begin{equation}
\rho_{\alpha} \sim ABCD. 
\label{rhoCTM}
\end{equation}
Since $\rho_{\alpha}$ is given by an infinite strip, one also needs infinite-size CTM's
in this relation. But exactly in this limit they are known for several non-critical
integrable models and have the form
\begin{equation}
A = e^{-u \,\mathcal{H}_{CTM}}  
\label{CTM}
\end{equation}
where $u$ contains the anisotropy of the two-dimensional system. This is a consequence 
of the star-triangle relations on which the integrability rests \cite{Cardy90}.
This approach gives $\mathcal{H}_{\alpha}$ in the original variables, see section 5.2, and
explicit expressions for the single-particle eigenvalues $\varepsilon_l$ in the 
diagonalized form. According to the derivation, it applies to one-half of an infinite
chain, but in practice the chain has only to be much longer than the correlation length. 
\par
Summing up, we have shown how to arrive at (\ref{rhogen}) and how to obtain the
$\varepsilon_l$. The eigenstates of $\rho_{\alpha}$ and their eigenvalues $w_n$ then
follow by specifying the occupation numbers of all single-particle levels. The
analytical result for $\varepsilon_l$ just mentioned is exceptional. For finite
subsystems beyond one or two sites, one has to find the $\varepsilon_l$ numerically.
This leads to a characteristic difficulty, because the eigenvalue equations in (II)
contain hyperbolic functions which approach $\pm 1$ for large $\varepsilon_l$. As the 
subsystem size grows, more and more values lie (exponentially) close to $\pm 1$, and
can only be obtained reliably with special techniques \cite{Chung/thesis}. Therefore 
the values of the $\varepsilon_l$ in most of the following figures do not exceed 20-30.

%
%--------------------------------------------------------------------------------
%
\section{Spectra}

In this section we give an overview of the single-particle spectra and the full 
$\rho_{\alpha}$-spectra for various situations. These include different dimensions,
critical and non-critical systems and the geometrical shape of the subsystem.
We will focus on the $\varepsilon_l$ because these are the primary quantities.

\subsection{One dimension}

(I) Non-critical chains.
\par
For infinite TI, XY and oscillator half-chains, the CTM 
approach gives the universal formula
\begin{equation}
 \varepsilon_l= \left\{ \begin{array}{r@{\quad,\quad}r}
(2l+1)\varepsilon & \mathrm{disordered} \; \mathrm{region} \\
2l\varepsilon & \quad \mathrm{ordered} \; \mathrm{region} 
\end{array} \right.  
\label{epsilonCTM}
\end{equation}
where $l=0,1,2,\dots$. Thus one has \emph{equidistant} levels and in a plot  
$\varepsilon_l \;vs.\; l$ the dispersion is strictly linear.
%%%%%%%%%%%%%%%%%%%%%%%%%%%%%%%%%%%%%%%%%%%%%%%%
\begin{figure}[htb]
\centering
\includegraphics[scale=.55]{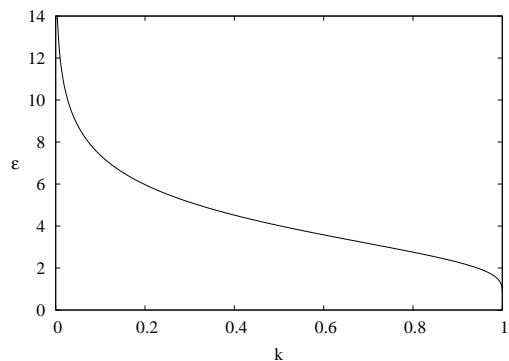}
\caption{Level spacing as a function of the parameter $k$.}
\label{fig:spacing}
\end{figure}
%%%%%%%%%%%%%%%%%%%%%%%%%%%%%%%%%%%%%%%%%%%%%%%%
The only free parameter
is the level spacing which depends on the details of the model. It is given by
\begin{equation}
\varepsilon=\pi\,I(k')/I(k),
\label{elliptic}
\end{equation}
where $I(k)$ denotes the complete elliptic integral of the first kind, and 
$k'= \sqrt{1-k^2}$.
The elliptic modulus $k$ with $0 \le k \le 1$  is given 
in the TI model by
\begin{equation}
 k = \left\{ \begin{array}{r@{\quad,\quad}r}
\lambda &  \lambda < 1 \\
1/\lambda & \lambda > 1 
\end{array} \right.  
\label{epsilonCTM2}
\end{equation}
In the XY model, the ordered region is subdivided by the so-called disorder line  
$\gamma^2+h^2 = 1$ and one has to distinguish three cases 
\begin{equation}
 k = \left\{ \begin{array}{r@{\quad,\quad}r}
\gamma/\sqrt{\gamma^2+h^2-1} &  h > 1   \\
\sqrt{\gamma^2+h^2-1}/\gamma &  \gamma^2+h^2 > 1, h < 1 \\
\sqrt{(1-\gamma^2-h^2)/(1-h^2)} &  \gamma^2+h^2 < 1, h < 1  
\end{array} \right.  
\label{epsilonCTM3}
\end{equation}
Here the last formula comes from a different approach \cite{Its05,Its/Korepin09}.
For the oscillator chain, $k$ is the nearest-neighbour coupling and one has to put 
$\omega_0=1-k$. In this case, there is no ordered region. In all models, the critical
point is given by $k=1$ and since $I(k)$ diverges for $k \rightarrow 1$, the level
spacing vanishes there and the dispersion curve becomes flat. The complete behaviour 
of $\varepsilon$ is shown in Fig. \ref{fig:spacing}.
\par
\begin{figure}[htb]
\centering
\includegraphics[scale=.43]{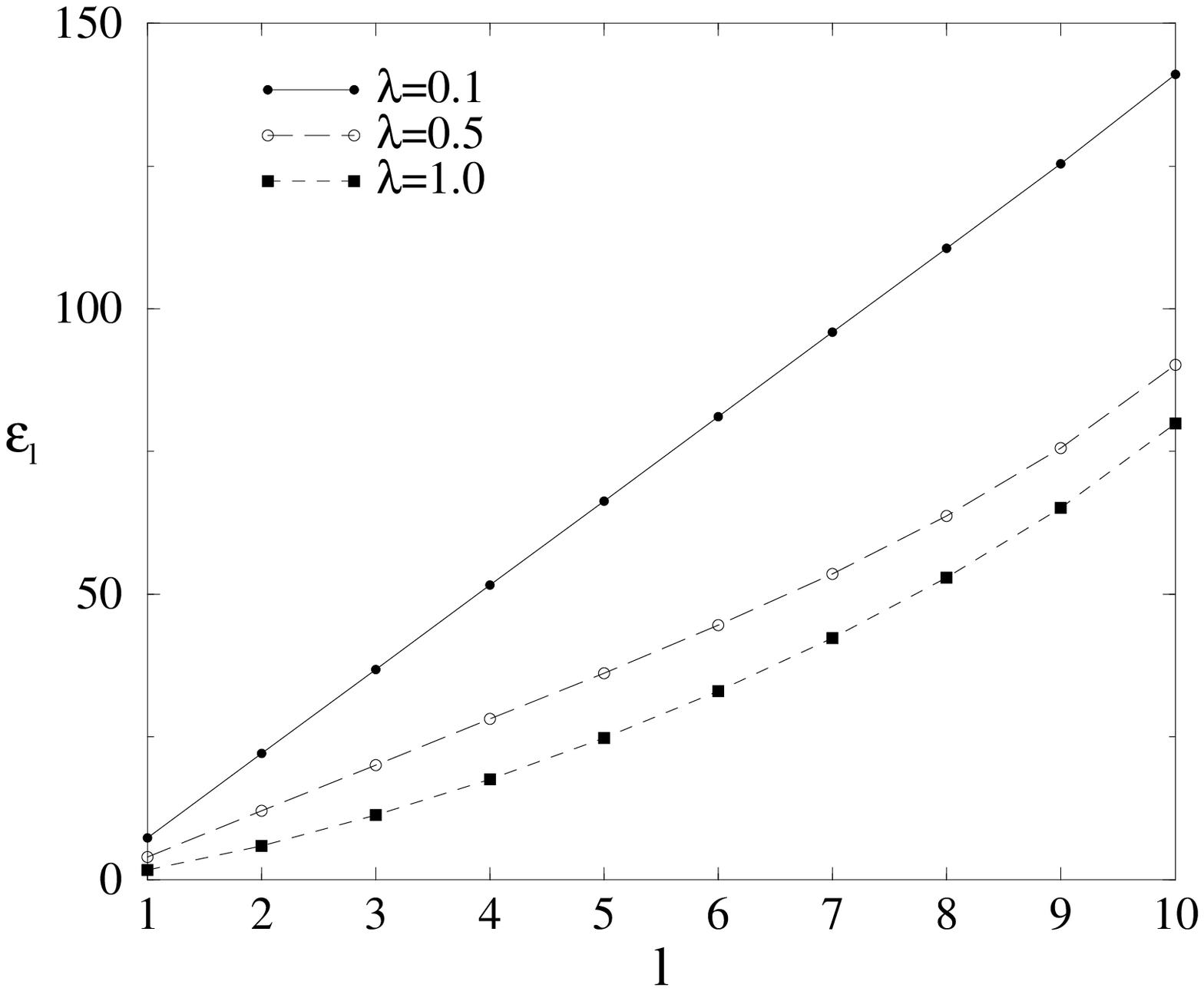}
\includegraphics[scale=.43]{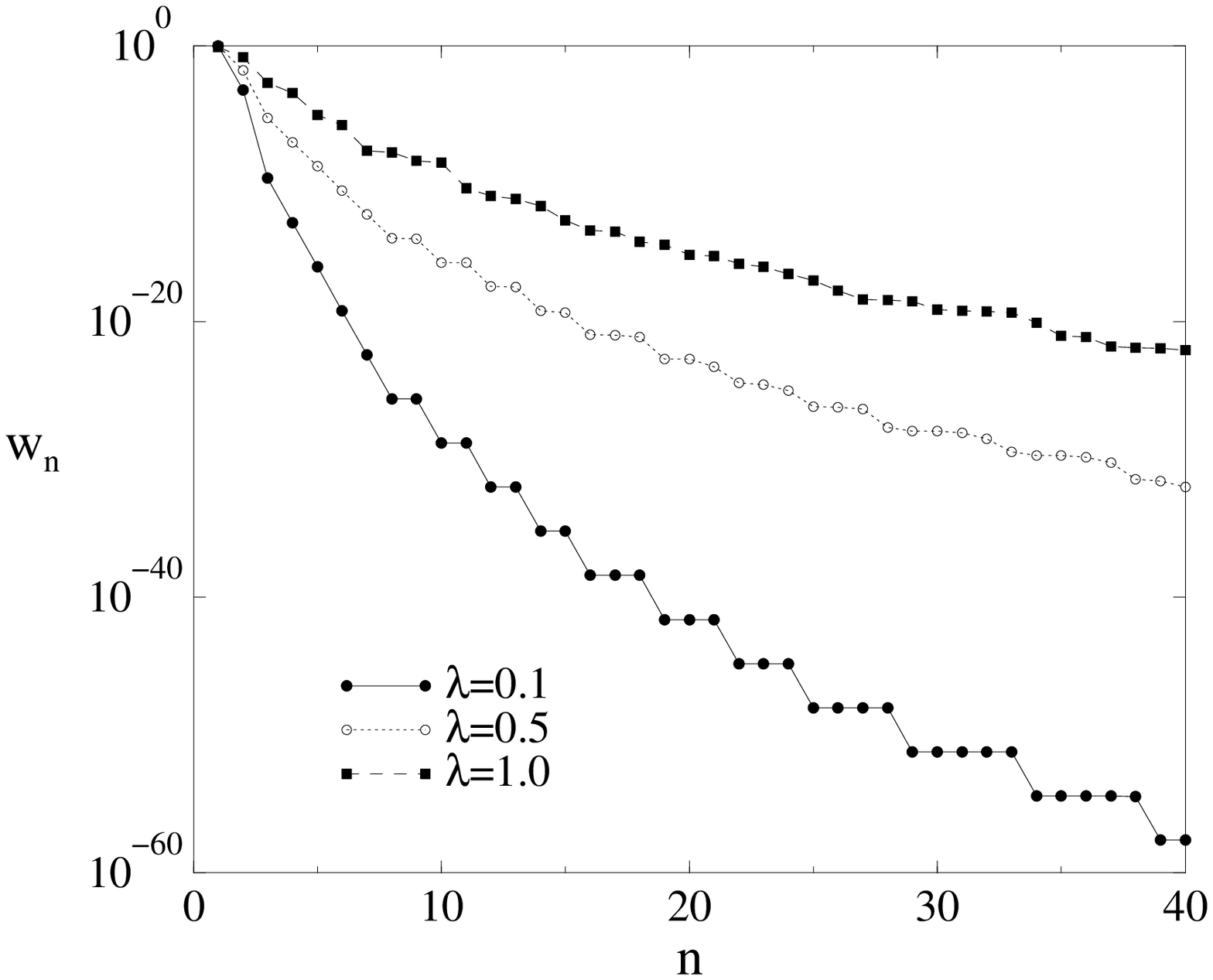}
\caption{Density-matrix spectra for one-half of a transverse Ising chain
with $N=20$ sites in its ground state. Left: All ten single-particle eigenvalues 
$\varepsilon_l$. Right: The largest total eigenvalues $w_n$. 
Reprinted with permission from \cite{Chung/Peschel01}. \copyright 2001 by the APS.}
\label{fig:spectra_Ising}
\end{figure}

Results for finite TI chains are shown in Fig. \ref{fig:spectra_Ising} on the left.
The linear behaviour is perfect for the smallest $\lambda$. As one comes closer to the
critical point, the slope decreases as predicted, but there are also deviations from
the linearity for large $\varepsilon_l$. Thus the linear region shrinks and is no longer
visible at the critical point. This is the typical finite-size scenario in these models.
On the right side, the resulting $w_n$, ordered by magnitude, are shown. One can see a 
rapid decrease with $n$ which is fastest for the smallest $\lambda$ but is still 
impressive at criticality (note the vertical scale). This means that a Schmidt 
decomposition could be truncated safely after about 10 terms and is the basis for the 
fantastic performance of the DMRG in this case \cite{Legeza/Fath96}.
\par
%%%%%%%%%%%%%%%%%%%%%%%%%%%%%%%%%%%%%%%%%%%%%%%%
%%%%%%%%%%%%%%%%%%%%%%%%%%%%%%%%%%%%%%%%%%%%%%%%
The lowest $w_n$-curve also shows a step structure with plateaus which become longer
with $n$. These are a consequence of the equidistant levels, a certain 
eigenvalue of $\mathcal{H}_{\alpha}$ can then be realized by different combinations of
$\varepsilon_l$. The degeneracy is given by the number of partitions $P(s)$ of an integer
$s$ into other (odd or even) integers. Using asymptotic formulae for the $P(s)$,
one finds the leading large-$n$ behaviour \cite{Okunishi/Hieida/Akutsu99} 
\begin{equation}
w_n \sim \exp[-a (\ln n)^2]
\label{asymptoticw}
\end{equation}
where $a=\varepsilon \,6/\pi^2$. The same result with a different constant $a$ holds for 
bosons. If the dispersion is not strictly linear, the steps are smeared and a rather
smooth $w_n$ spectrum is obtained.  
\par
An important new feature appears in the $\varepsilon_l$-spectra, if the subsystem 
is a segment in a chain. Then a two-fold degeneracy is found, at least for the lowest
eigenvalues. The reason lies in the form of the eigenfunctions, which are concentrated
near the ends, as will be demonstrated in section 5. This leads to a degeneracy of the 
$w_n$, with a factor of 2 for each $\varepsilon_l$ which is involved, and therefore to a
significantly slower decay.
\par
For the spin chains, there are cases where the ground state simplifies and becomes a 
doublet of product states. Then one $\varepsilon_l$ is zero, while all others diverge.
As a consequence, all $w_n$ except two collapse to zero.
This happens not only in the TI model for $\lambda \rightarrow \infty$, but also in the
XY model on the disorder line \cite{Chung/Peschel01}. If the result were not known,
one could locate the line from the behaviour of the spectra.
\par
Finally, we note that also a dimerized half-filled hopping model shows such equidistant
$\varepsilon_l$ because one can relate it to the TI model via the correlation functions.
The parameter $k$ is then given by $k=(1-\delta)/(1+\delta)$, where $\delta > 0$ is the
dimerization parameter.
\bigskip\par
\noindent (II) Critical chains.
\par
\begin{figure}[htb]
\centering
\includegraphics[scale=.6]{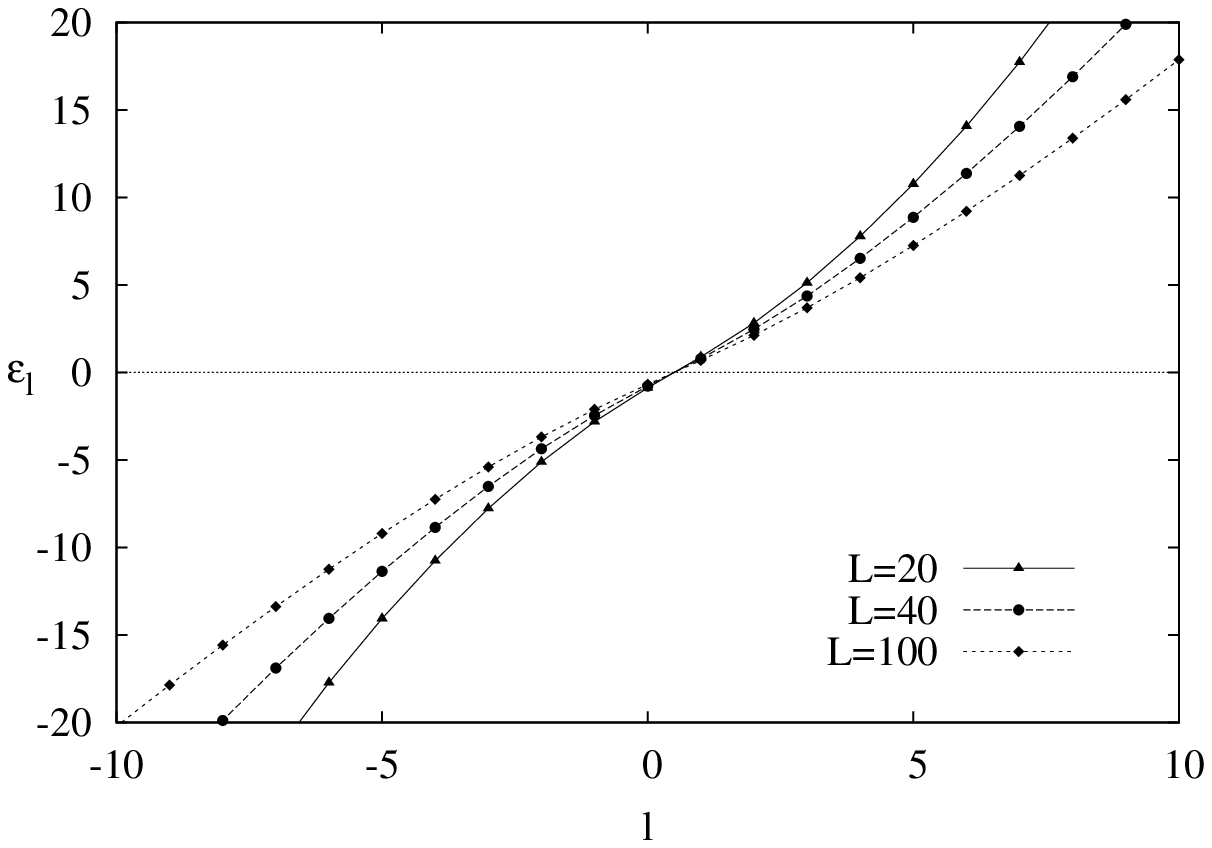}
\includegraphics[scale=.6]{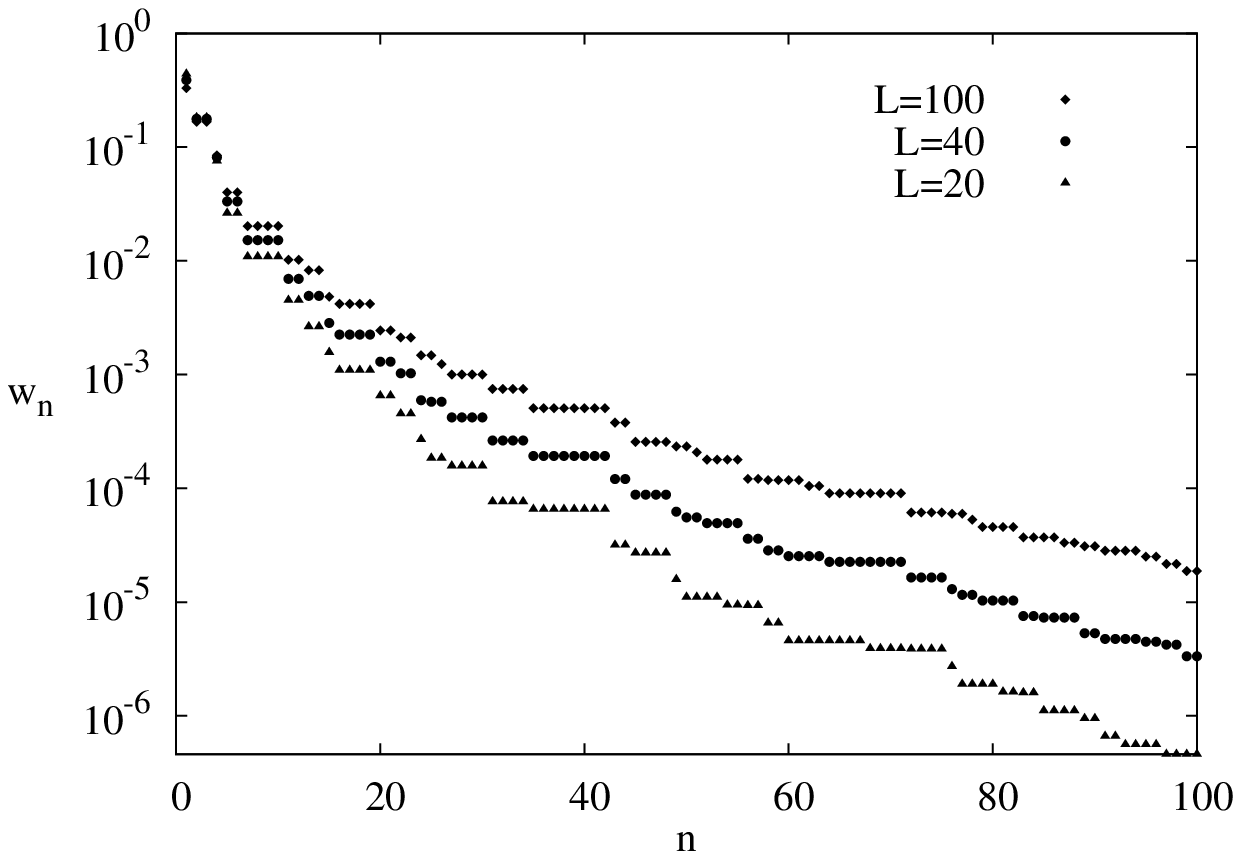}
\caption{Size dependence of the density-matrix spectra in a critical system.
Shown are results for segments of different lengths in an infinite hopping model.
Left: Single-particle eigenvalues $\varepsilon_l$. Right: Total eigenvalues $w_n$.
After Ref. \cite{Greifswald08}.}
\label{fig:spectra_XX}
\end{figure}

In critical systems, the size of the subsystem affects not only the upper part of the 
single-particle spectrum. This is shown in Fig. \ref{fig:spectra_XX} for a segment in a 
half-filled hopping model, or XX chain.
%%%%%%%%%%%%%%%%%%%%%%%%%%%%%%%%%%%%%%%%%%%%%%%%
%%%%%%%%%%%%%%%%%%%%%%%%%%%%%%%%%%%%%%%%%%%%%%%%
The eigenvalues follow in this case from the simple correlation
matrix
\begin{equation}
C_{m,n}= \int_{-k_F}^{k_F} \frac {\mathrm{d}q}{2\pi} \, e^{-iq(m-n)} =
\frac {\sin(k_F(m-n))}{\pi(m-n)}
\label{corrXX}
\end{equation}
where $k_F=\pi/2$ for half filling.
One sees that the whole dispersion curve is shifted towards the horizontal axis 
and becomes flatter as the length increases. The shift is not rapid, the first few 
eigenvalues vary as  $1/(\ln L + b)$ with somewhat different constants $b$ around 2.5. 
From a continuum approximation for the eigenvalue problem, one obtains the asymptotic 
formula
\begin{equation}
 \varepsilon_l = \pm \;\frac{\pi^2}{2\ln L} (2l-1) \;, \;\;\; l = 1,2,3\dots 
\label{epsilonconf}
\end{equation}
which can also be derived with conformal considerations \cite{Peschel04}. A similar
expression for bosons was given in \cite{Callan/Wilczek94}. The formula is also valid 
for a segment of $L$ sites at the end of a chain, if one substitutes 
$2\ln L \rightarrow  \ln(2L)$,  which increases the values roughly by 2. 
It predicts the $1/\ln L$ behaviour, but also a linear dispersion as in the non-critical 
case. In practice, this can only be seen if in addition to $L$ also $\ln L$ is large,
which requires huge sizes.
Nevertheless, it is an important guide for the understanding of the situation and will 
be used again later. Formulae of this type and the numerical difficulties in verifying 
them are known from studies of critical finite-size CTM's 
\cite{Truong/Peschel88,Davies89,Davies/Pearce90}. 
\par
Although the change of the $\varepsilon_l$ is slow, it has a clear effect on the 
$w_n$ spectra, as seen on the right of the figure. The decay becomes significantly
slower for larger systems, which means that the entanglement grows with the size.
Invoking conformal results, one can obtain the functional form of the $w_n$-spectrum
\cite{Calabrese/Lefevre08,CCreview}.
Asymptotically, (\ref{asymptoticw}) is still valid, but now $a \sim 1/\ln L$ varies 
with the length.
Therefore the DMRG method does not work as well in this case, although it still can
handle sizes of $L \sim 100 $.
\par
\begin{figure}[htb]
\centering
\includegraphics[scale=.55]{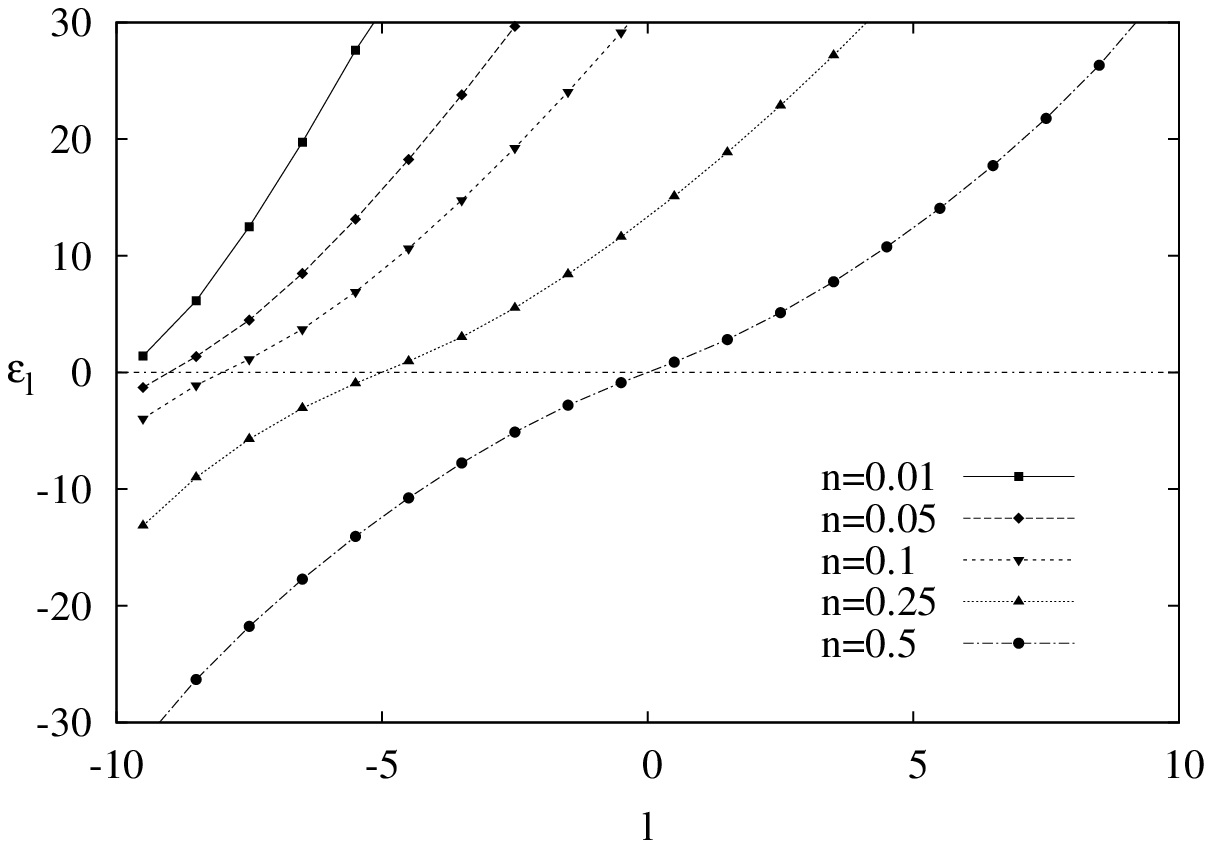}
\includegraphics[scale=.55]{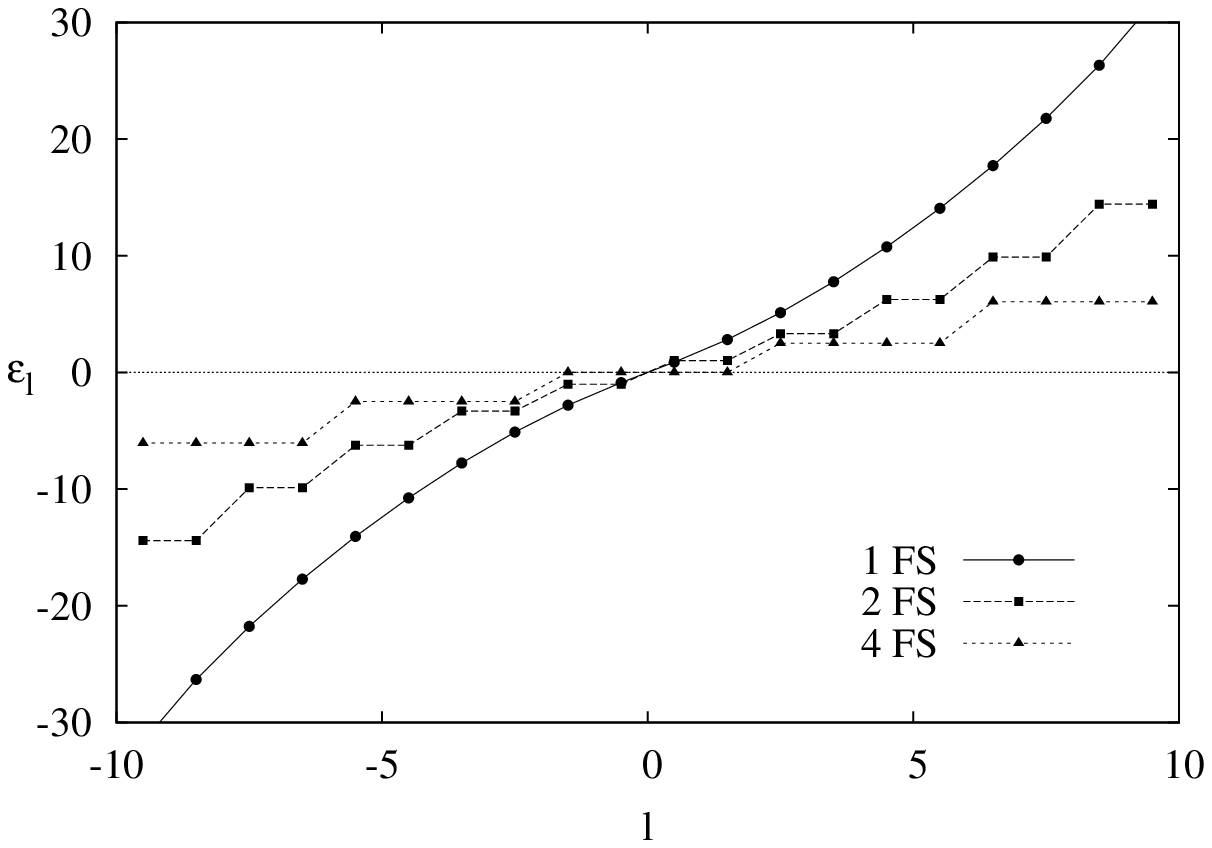}
\caption{Single-particle spectra for different ground states. Left: Variation with the
filling. Right: Variation with the number of equal-size Fermi seas at half filling.
All results are for a segment of $L= 20$ sites in an infinite hopping model.}
\label{fig:fill_fermisee}
\end{figure}

Finally we want to show how certain modifications of the ground state affect the 
spectra.
In the previous cases, the system was always half filled, which leads to a symmetric
spectrum ($\pm \varepsilon_l$ appear) \cite{Cheong/Henley04/1}. If the filling is varied in 
(\ref{corrXX}), one finds 
that the $\varepsilon_l$-dispersion curve is moved up or down in a similar way as the
Fermi level, see Fig. \ref{fig:fill_fermisee}. For a completely full or empty system,
which is a product state (in spin language all spins are up or down), the $\varepsilon_l$ 
are all infinite and $w_n$ becomes a Kronecker symbol, $w_n=\delta_{n,1}$, as it should.
\par
%%%%%%%%%%%%%%%%%%%%%%%%%%%%%%%%%%%%%%%%%%%%%%%%
%%%%%%%%%%%%%%%%%%%%%%%%%%%%%%%%%%%%%%%%%%%%%%%%
%
If the Fermi sea consists of several disconnected parts, one finds degeneracies in
the eigenvalues, if empty and full regions in momentum space have equal size. 
This is shown in Fig. \ref{fig:fill_fermisee} on the right. It looks as if
one had several independent kinds of particles. Effectively, the dispersion then rises 
only with a fraction of the slope. The same holds in the case of non-equal Fermi seas, 
where the degeneracies are washed out.
Such a situation occurs, for example, for the ground state of the chain with an energy 
current \cite{Eisler/Zimboras05}. As in the previous examples, the $w_n$  then decrease 
more slowly and the entanglement becomes larger.
\par
\begin{figure}[htb]
\centering
\includegraphics*[scale=.28,angle=270]{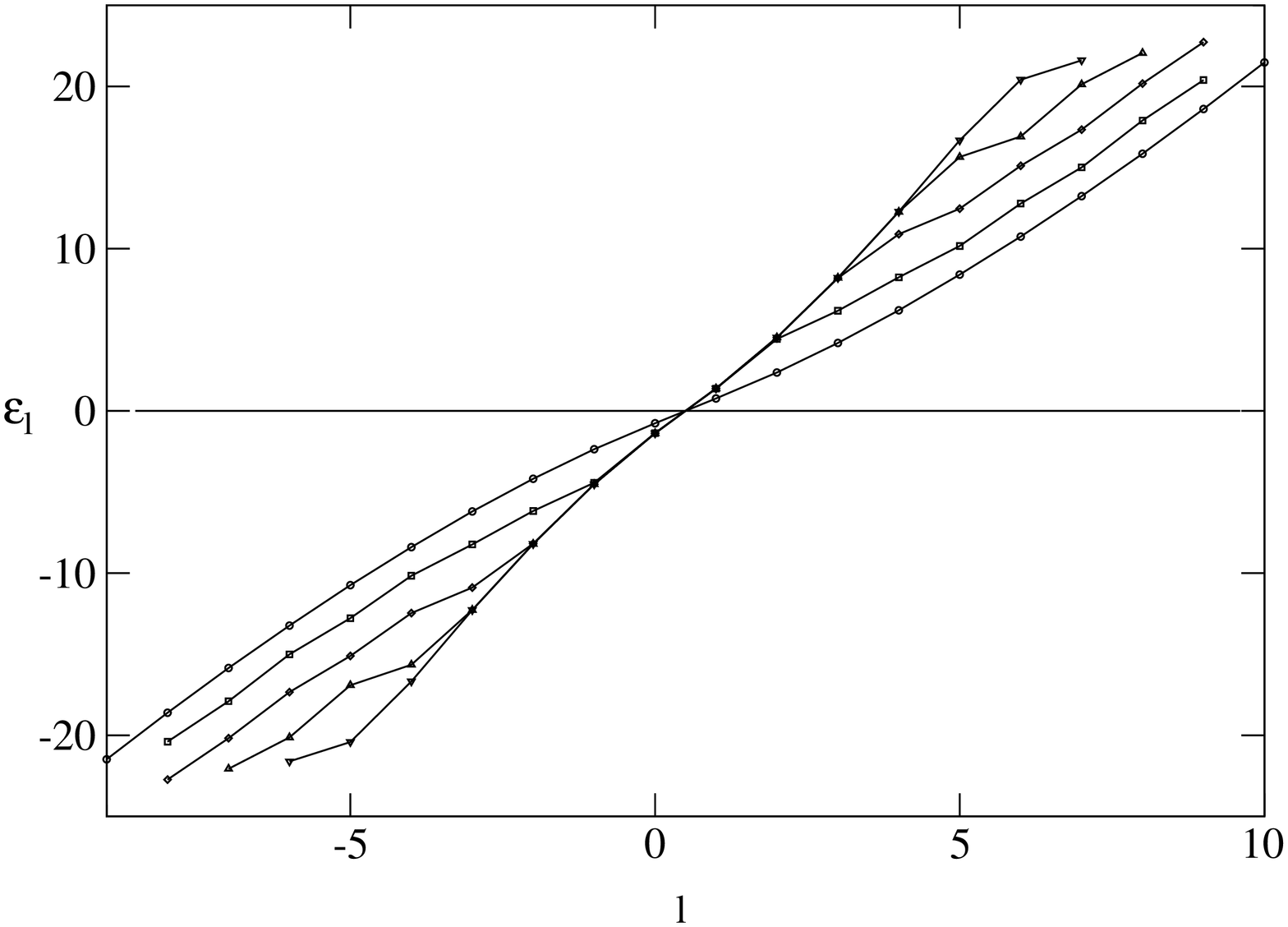}
\includegraphics*[scale=.28,angle=270]{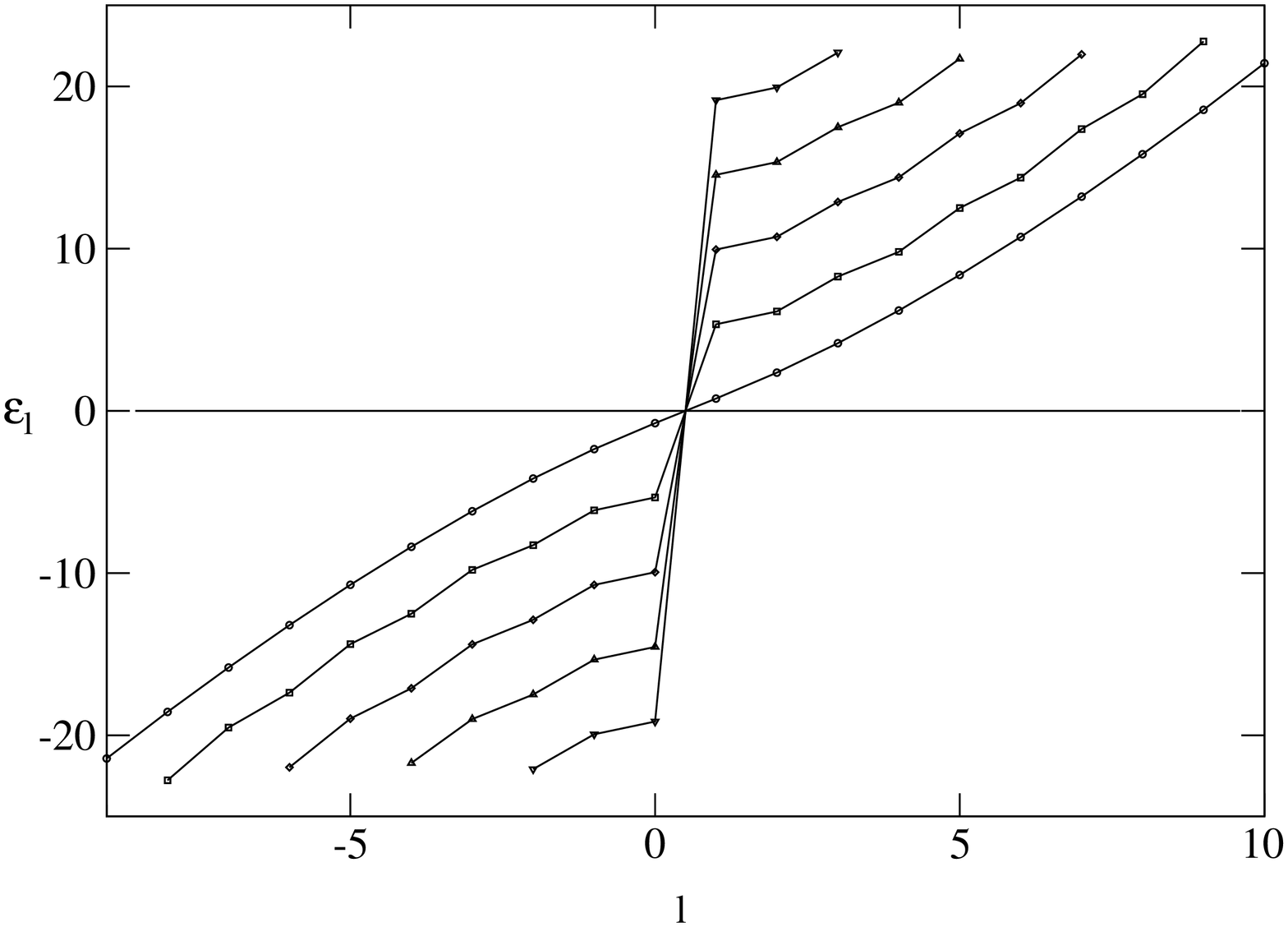}
\caption{Influence of interface modifications on the single-particle spectra for a segment 
with $L=50$ sites in a hopping model.
Left: Modified bond at one end. Right: Modified bonds at both ends.
The curves correspond to bond values $t=1; 10^{-1}; 10^{-2}; 10^{-3}; 10^{-4}$, from bottom
to top in the right part of the figures. After Ref. \cite{Peschel05}. }
\label{fig:eps_defect}
\end{figure}

If one modifies the hopping between the segment and the environment at one interface,
one can interpolate continuously between a homogeneous chain and one with an open end. 
\cite{Peschel05}. The $\varepsilon_l$-spectrum in this case is shown in Fig. 
\ref{fig:eps_defect} on the left.
As the bond is weakened, a region with a steeper initial ascent appears before the curve 
follows the pattern without defect. This region can be associated with the developing free 
end and remains when the bond is cut completely. If, on the other hand, the bonds at both
interfaces are weakened, the dispersion is shifted upwards resp. downwards as a whole and
a gap develops. In the decoupling limit it goes to infinity and the entanglement vanishes.
%%%%%%%%%%%%%%%%%%%%%%%%%%%%%%%%%%%%%%%%%%%%%%%%
%%%%%%%%%%%%%%%%%%%%%%%%%%%%%%%%%%%%%%%%%%%%%%%%

\subsection{Two dimensions}

(I) Non-critical systems
\par
The simplest two-dimensional system consists of a set of $M$ uncoupled identical
parallel chains, all divided at the same point such that the subsystem has the form
of a half-strip \cite{Croo99}. This is the usual DMRG geometry. The combined RDM is 
then a product of the individual ones and ${\mathcal{H}_{\alpha}}$ becomes a sum 
\begin{equation}
{\mathcal{H}_{\alpha}} = \sum_{l,\mu} \varepsilon_{l,\mu}\, f_{l,\mu}^{\dagger} f_{l,\mu} 
\label{rhochains}
\end{equation}
where $\mu$ is the chain index. Since  $\varepsilon_{l,\mu}= \varepsilon_l$, the 
single-particle eigenvalues are simply $M$-fold degenerate. For free particles,
a coupling of the chains does not change this situation because one can separate the
system into $M$ new independent chains by a Fourier transformation in the perpendicular 
direction \cite{Chung/Peschel00,Cramer/Eisert/Plenio07}. The index $\mu$ in 
(\ref{rhochains}) then becomes the Fourier index $q$. Only $\varepsilon_{l,q}$ will 
depend on $q$ and the $M$-fold degenerate levels will become bands. 
\par
\begin{figure}[htb]
\centering
\includegraphics[scale=.5]{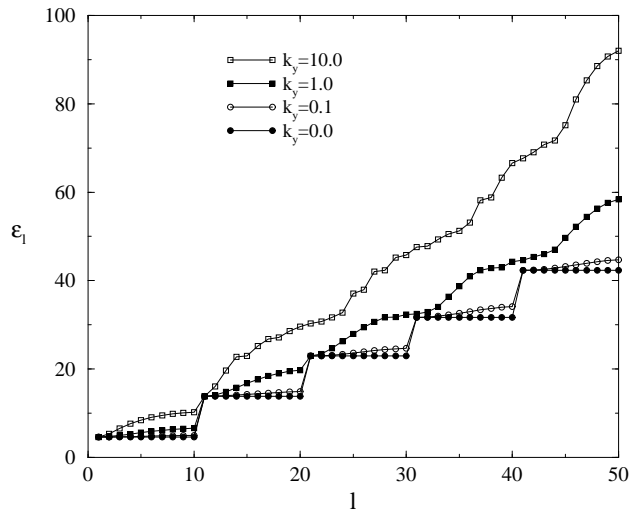}
\caption{Single-particle eigenvalues for one-half of a $10 \times 10$ system of coupled
oscillators with $\omega_0=k_x=1$ and different couplings $k_y$.
Reprinted with permission from \cite{Chung/Peschel00}. \copyright 2000 by the APS.}
\label{fig:eps_osc2d}
\end{figure}

For coupled oscillators and an infinite half-strip, the problem can in this way be solved 
\emph{exactly} by invoking the one-dimensional results. One only has to determine the 
elliptic parameter $k=k(q)$  for each Fourier component from the coupling 
$k_x$ in the chain direction and the frequency $\omega^2(q)= \omega^2_0+2k_y(1-\cos q)$
via 
\begin{equation}
\frac {\omega(q)}{k_x} = \frac {1-k}{k}
\label{elliptic2}
\end{equation}
Numerical results for a system of 10 chains with actually finite length are shown in
Fig. \ref{fig:eps_osc2d}. 
%%%%%%%%%%%%%%%%%%%%%%%%%%%%%%%%%%%%%%%%%%%%%%%%
%%%%%%%%%%%%%%%%%%%%%%%%%%%%%%%%%%%%%%%%%%%%%%%%
The coupling of the chains was varied and one can see nicely, how the
plateaus with 10 levels develop into bands and a rather smooth, roughly linear curve 
results in the isotropic case. The initial plateaus, combined with the large freedom
in the bosonic occupation numbers, lead to even larger plateaus in the $w_n$-spectrum.
In the isotropic case, one can derive an asymptotic formula as 
(\ref{asymptoticw}) by assuming a strictly linear behaviour with a slope 
\begin{equation}
\varepsilon_l = \lambda\,l = \frac{\varepsilon}{M} \, l
\label{epseffective}
\end{equation}
Then one finds (\ref{asymptoticw}) with a coefficient  
$a= \lambda \, 3/ 2 \pi^2$. The crucial difference is that $\lambda  \sim 1/M$ depends
inversely on the width, which makes the decay of the $w_n$ exceedingly slow for
wide systems. The entanglement becomes correspondingly high. This is a general feature
and will be taken up again in section 6. For the DMRG it means that the width of the
strip puts a fundamental limit on its applicability.
\par
For subsystems in the form of $L \times L$ squares embedded in an infinite lattice, one 
can obtain similar results by solving the equation (\ref{boson}) numerically.
One finds again bands as in Fig. \ref{fig:eps_osc2d}, but the number of 
states in the lowest bands is now given by $(4L-4)$, which one recognizes as the
the number of boundary sites. Plotting the $\varepsilon_l$ as a function of the scaled 
index $l/(4L-4)$, the results fall essentially on top of each other. This is the same 
behaviour as for the single straight boundary, where $l/M$ enters. It is a clear indication 
that the single-particle states are associated with the interface between the subsystem 
and its surrounding, as in one dimension. 
\par
\noindent (II) Critical systems
\par
In this case one finds similar features which we will exhibit for the hopping model
on a square lattice. The isotropic half-filled model has the well-known quadratic
Fermi surface with corners at $(\pm \pi,0)$ and  $(0,\pm \pi)$ in momentum space. This
gives the correlation function as the product of two one-dimensional ones as in
(\ref{corrXX}) 
\begin{equation}
C(x,y|0,0) = 2 \frac {\sin(\pi(x-y)/2)}{\pi(x-y)}
\frac {\sin(\pi(x+y)/2)}{\pi(x+y)}
\label{corrXX2d}
\end{equation}
where $x$ and $y$ are integers. If the model is anisotropic, the Fermi surface is more
complicated and one momentum integration has to be done numerically. With these functions
one can calculate the spectra for arbitrary subsystems embedded in an infinite 
lattice. For half filling, the spectra are again symmetric, i.e. the eigenvalues occur
in pairs $\pm \varepsilon$.
\par
\begin{figure}[htb]
\centering
\includegraphics[scale=.57]{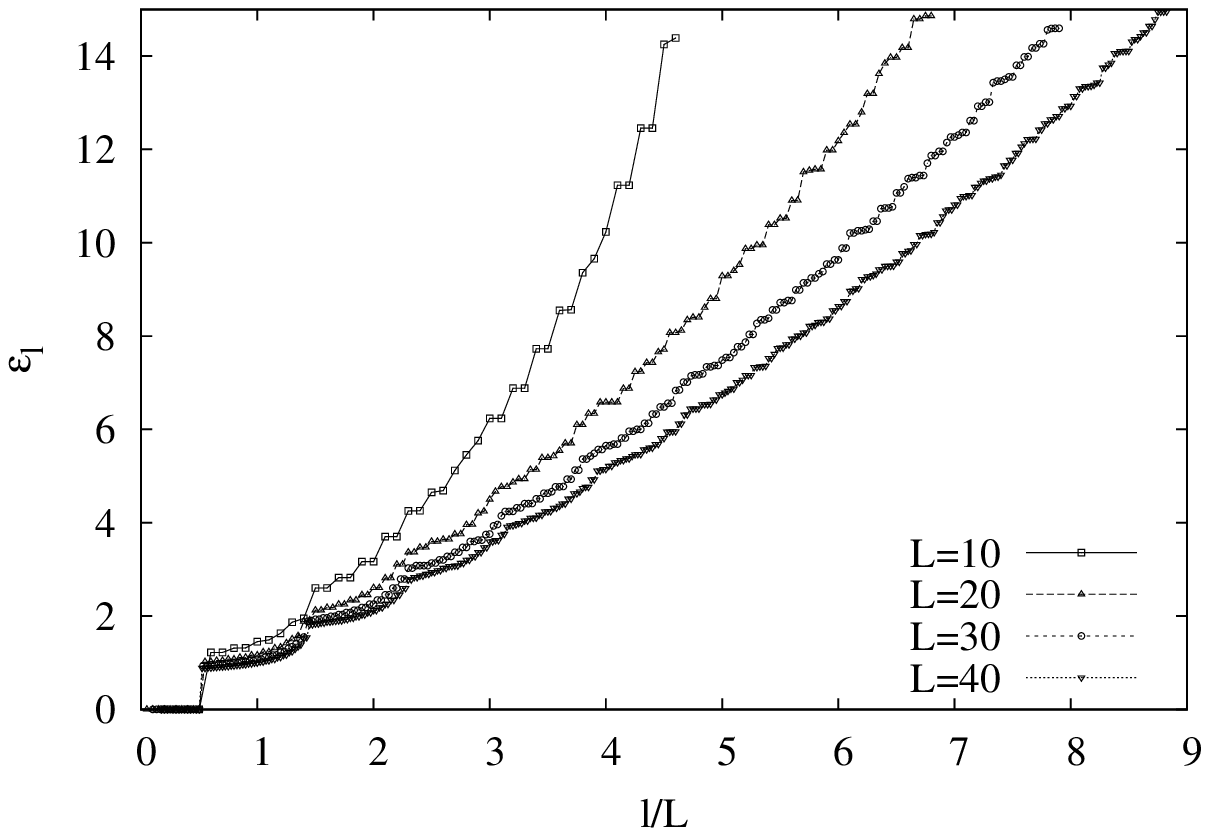}
\includegraphics[scale=.57]{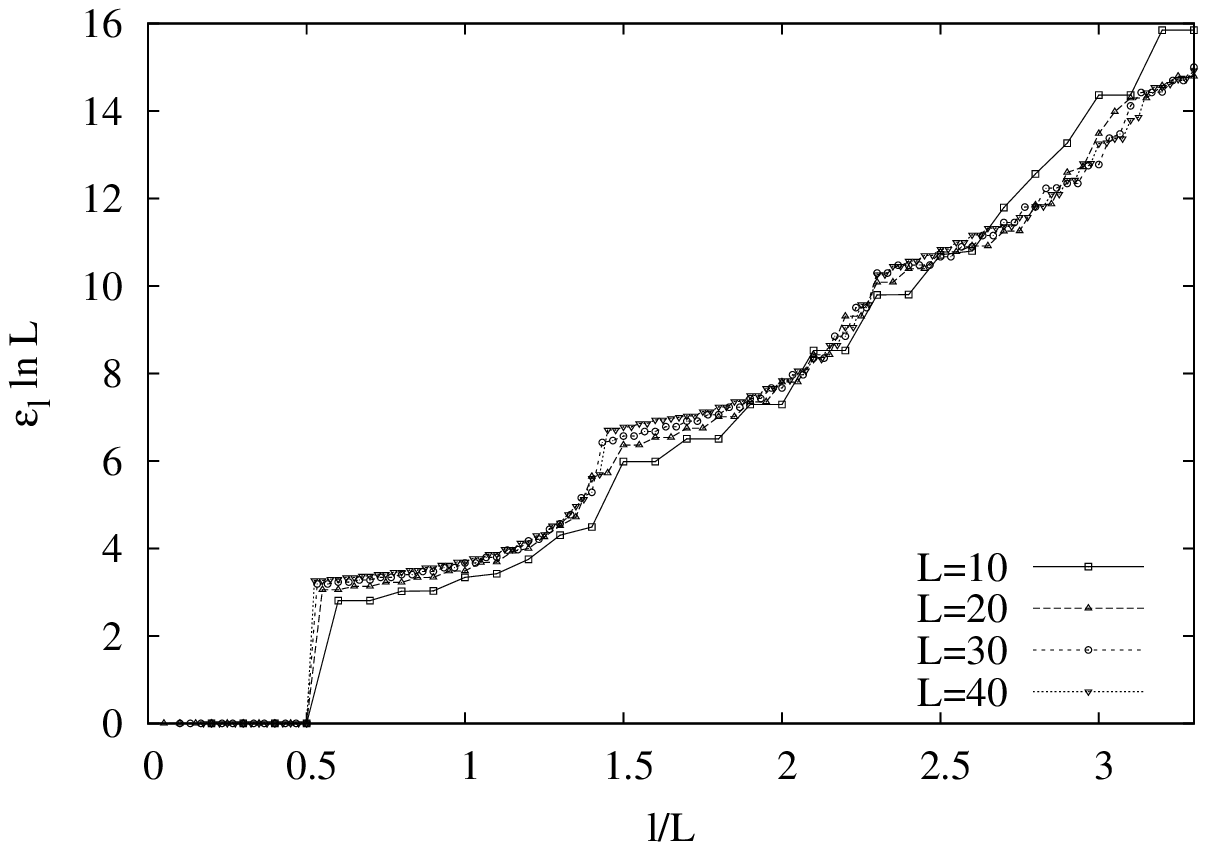}
\caption{Single-particle spectra for $L \times L$ squares in an infinite planar hopping
model. Left: $\varepsilon_l$ vs. $l/L$. Right: $\varepsilon_l \ln L$ vs. $l/L$. Only the 
positive eigenvalues are shown.}
\label{fig:eps_xx2d}
\end{figure}

Fig. \ref{fig:eps_xx2d} shows results for $L \times L$ squares, plotted to exhibit the
scaling behaviour. On the left, the $\varepsilon_l$ are shown as a function of the 
scaled index $l/L$. One can see low-lying bands which all have the same horizontal length 1
and thus contain $L$ states. However, their height still varies with $L$. Only by plotting
$\varepsilon_l \ln L$ they all collapse on one curve, as shown on the right. This demonstrates
that, on the one hand, the linear size $L$ enters as in the non-critical case, but that also
the inverse logarithmic dependence on $L$ found in one dimension remains.
As a result, logarithmic corrections appear in the entanglement entropy, see section 6.
Note also that $L$ enters, and not $4L-4$ as before. This is most obvious in a band of $L$ eigenvalues
which are exactly zero (the figure shows only one-half of it). The latter feature is peculiar 
to the square and does not occur for rectangles, where the dispersion rises smoothly from 
zero. 

%%%%%%%%%%%%%%%%%%%%%%%%%%%%%%%%%%%%%%%%%%%%%%%%
\begin{figure}[htb]
\centering
\includegraphics[scale=.7]{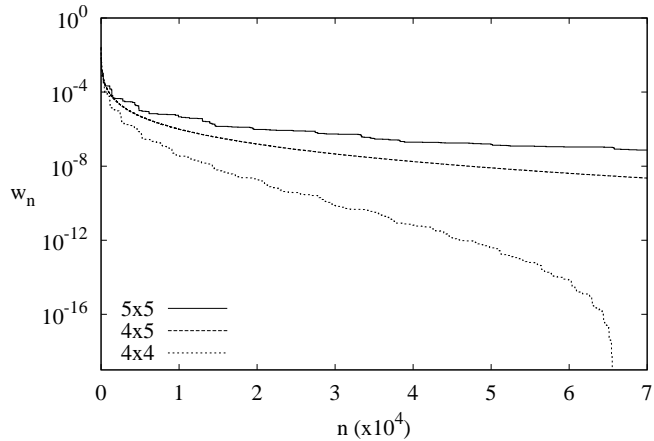}
\caption{Total eigenvalues $w_n$ for squares and rectangles in an infinite planar hopping
model. For the $4 \times 4$ system, the figure shows \emph{all}  $w_n$.}
\label{fig:wn_xx}
\end{figure}

%%%%%%%%%%%%%%%%%%%%%%%%%%%%%%%%%%%%%%%%%%%%%%%%
The resulting spectrum of $\rho_{\alpha}$ is shown in Fig. \ref{fig:wn_xx} for three
relatively small systems. For the  $4 \times 4$ square, all $2^{16}$ eigenvalues are
displayed and the s-shaped curve actually reflects the symmetry of the $\varepsilon_l$
spectrum. The $4 \times 5$ system gives much smoother results which can be fitted well by the
law (\ref{asymptoticw}). For it, and also for the $5 \times 5$ system, the curves drop only
to a value of about $10^{-4}$ for $n$ around $1000$, which is to be compared with the 
one-dimensional results of Fig. \ref{fig:spectra_XX}, where this value is reached already at
$n \sim 100$ for $L=100$. The same feature is found for other geometries 
\cite{Chung/Peschel01,Chung/thesis}. 
This shows very clearly the basic difference between one and two
(and also higher) dimensions.
%%%%%%%%%%%%%%%%%%%%%%%%%%%%%%%%%%%%%%%%%%%%%%%%
%%%%%%%%%%%%%%%%%%%%%%%%%%%%%%%%%%%%%%%%%%%%%%%%

\section{Further aspects}

\subsection{Single-particle wave functions}

(I) Chains
\par
\begin{figure}[thb]
\centering
\includegraphics[scale=.27,angle=270]{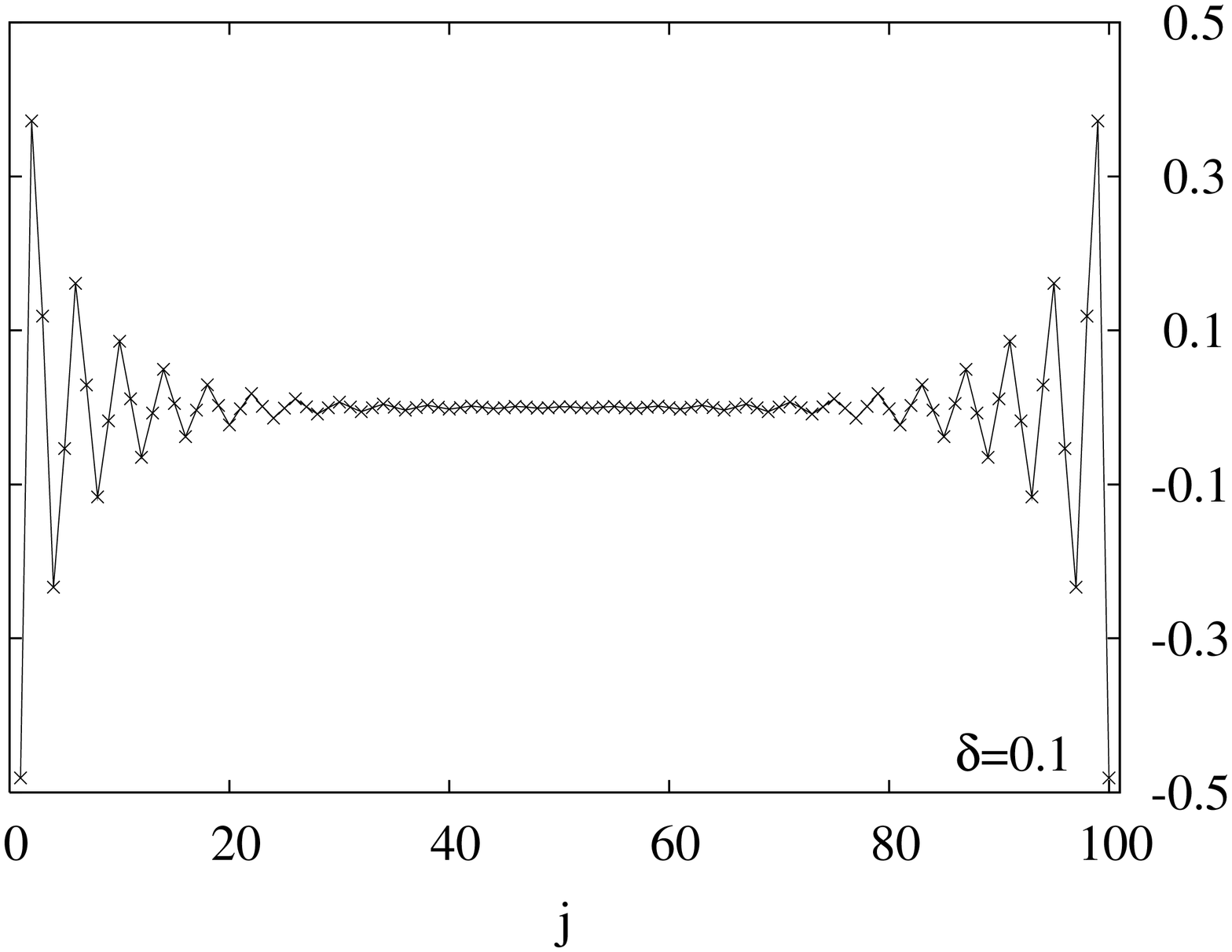}
\includegraphics[scale=.27,angle=270]{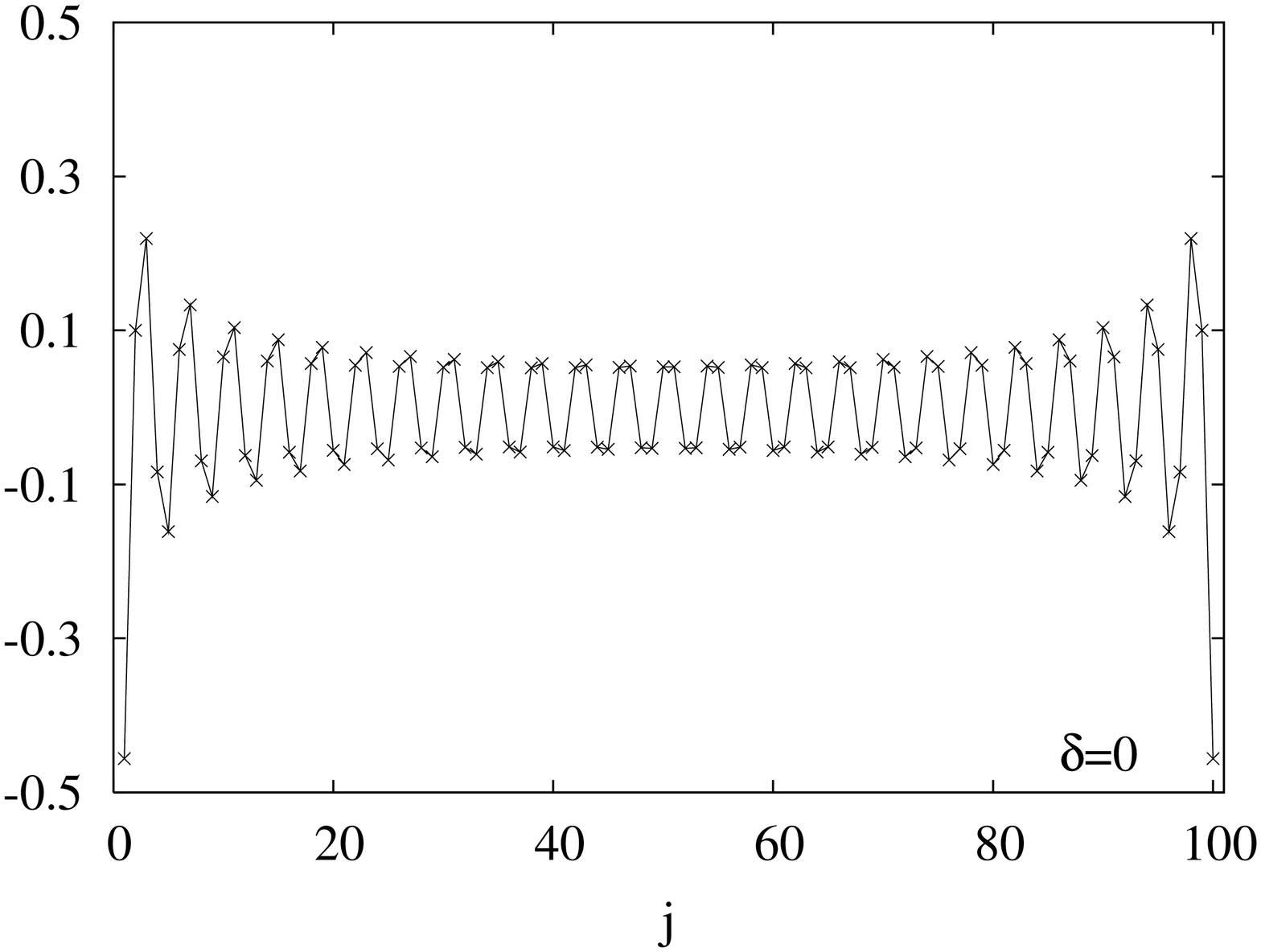}
\caption{Lowest lying single-particle eigenstates in a dimerized ($\delta=0.1$, left) 
and a homogeneous ($\delta=0$, right) hopping model for a segment of $L=100$ sites.
After Ref. \cite{Greifswald08}.}
\label{fig:eigvecs}
\end{figure}

The eigenfunctions associated with the $\varepsilon_l$ have a particular nature. 
In Fig.\ref{fig:eigvecs} they are shown for the smallest $\varepsilon_l$ in the
case of a segment in a half-filled hopping model. On the left, the model is dimerized,
i.e. non-critical, and one sees that the amplitude is concentrated near the two
interfaces to the remainder and almost zero in the middle. This feature persists
even in the homogeneous critical case seen on the right, although there is now a
slow decay into the interior. For the highest $\varepsilon_l$, on the other hand,
the amplitude is concentrated in the centre of the subsystem and the eigenfunction 
resembles a Gaussian. The same pattern can be seen in oscillator chains 
\cite{Chung/Peschel00,Gaite01,Botero/Reznik04}. It is very suggestive, since it means that 
the states which are most important in the entanglement are those closest to the boundary. 
The whole entanglement appears as a phenomenon taking place within a layer whose width is 
given by the correlation length.
\par
%
%%%%%%%%%%%%%%%%%%%%%%%%%%%%%%%%%%%%%%%%%%%%%%%%
%%%%%%%%%%%%%%%%%%%%%%%%%%%%%%%%%%%%%%%%%%%%%%%%
%
A lattice result for $\phi_l(j)$ in the hopping model is only available in the case 
$\varepsilon_0=0$ which occurs for odd $L$ \cite{Peschel04}. The wave function then is 
u-shaped and vanishes at every second site, see \cite{Cheong/Henley04/2}. However, one can 
derive an expression in the continuum limit. Putting $x=j/L$, it reads for a segment located
between $x=0$ and $x=1$  \cite{Peschel04}
\begin{equation}
\phi_l(x)= \frac{c}{\sqrt{x(1-x)}} \sin[\,\frac{\varepsilon_l}{2}\,\ln{(\frac{x}{1-x})}+\alpha]
\label{continphi}
\end{equation}
Such logarithmic oscillations were found earlier in CTM studies related to the TI 
\cite{Davies/Pearce90} and the oscillator half-chain \cite{Peschel/Truong91}.
In the latter case, which was also treated in \cite{Callan/Wilczek94},  
the square-root prefactor is absent.

\bigskip\par
\noindent (II) Planar systems
\par
\begin{figure}[htb]
\centering
\includegraphics[scale=.85]{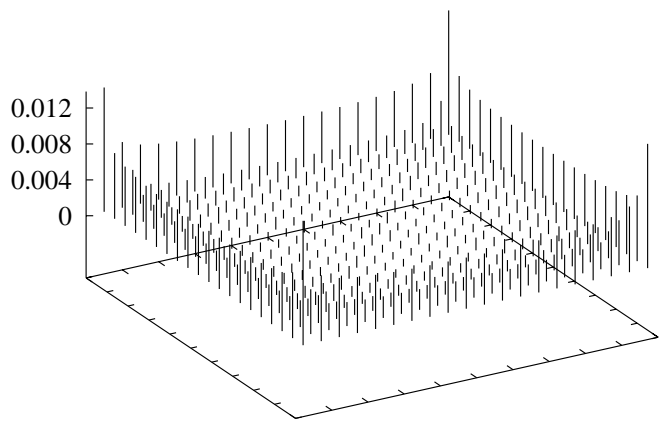}
\caption{Squared amplitudes summed over the band of states with $\varepsilon_l=0$
for a $20 \times 20$ square in an infinite planar hopping model.}
\label{fig:evec_2d}
\end{figure}

It is clear that the basic feature, namely the concentration near the interface,
will also be found in two dimensions. As an example, Fig. \ref{fig:evec_2d} shows the  
%%%%%%%%%%%%%%%%%%%%%%%%%%%%%%%%%%%%%%%%%%%%%%%%
%%%%%%%%%%%%%%%%%%%%%%%%%%%%%%%%%%%%%%%%%%%%%%%%
situation for the $\varepsilon_l=0$ states which occur for a quadratic subsystem in a planar
hopping model. Due to the degeneracy, the individual states are not uniquely defined and 
one has to consider all simultaneously. The maxima at the boundary are clearly visible and
one has the same u-shaped pattern as in the one-dimensional case. In addition, there is a 
slight enhancement along the diagonals. In general, the eigenfunctions have variations
parallel to the interface which are related to the square symmetry. To bring out their 
``radial''behaviour one has to calculate the analogue of a radial distribution function. One 
then sees, apart from a small bump in the centre, a clear increase towards the boundary for 
all low-lying bands. This will be even more pronounced in a non-critical system.

\subsection{Nature of $\mathcal{H}_{\alpha}$}

The eigenfunctions presented in the previous subsection have their origin in a
particular form of the effective Hamiltonian, which we now address.
In the CTM approach it is possible to give an explicit expression for $\mathcal{H}_{\alpha}$. 
This is done by considering (\ref{CTM}) in the limit of a very anisotropic system
\cite{Baxter77,Davies88,Truong/Peschel89}. 
For the TI half-chain this leads to the result
\begin{equation}
{\mathcal{H}}_{\alpha} = - C \left[ \sum_{n\ge 1} (2n-1)\sigma^z_n + 
\lambda \sum_{n\ge 1} 2n \sigma^x_n \sigma^x_{n+1} \right]
\label{hamCTM}
\end{equation}
where the constant $C$ depends on $\lambda$. Therefore $\mathcal{H}_{\alpha}$ also
describes a TI chain, but an inhomogeneous one, with coefficients which increase 
linearly away from the interface. In the two-dimensional problem, this reflects the
wedge-shaped geometry. In the RDM context, it suppresses the influence of sites far in
the interior because $\mathcal{H}_{\alpha}$ enters exponentially into $\rho_{\alpha}$.
This Hamiltonian can also be diagonalized directly \cite{Davies88,Truong/Peschel89} 
and one recovers the result (\ref{epsilonCTM}) for $\varepsilon_l$. In the limits
$\lambda \rightarrow 0$ and $\lambda \rightarrow \infty$, the level structure
(\ref{epsilonCTM}) can directly be read off the coefficients in (\ref{hamCTM}).
\par
\begin{figure}[htb]
\centering
\includegraphics[scale=.5]{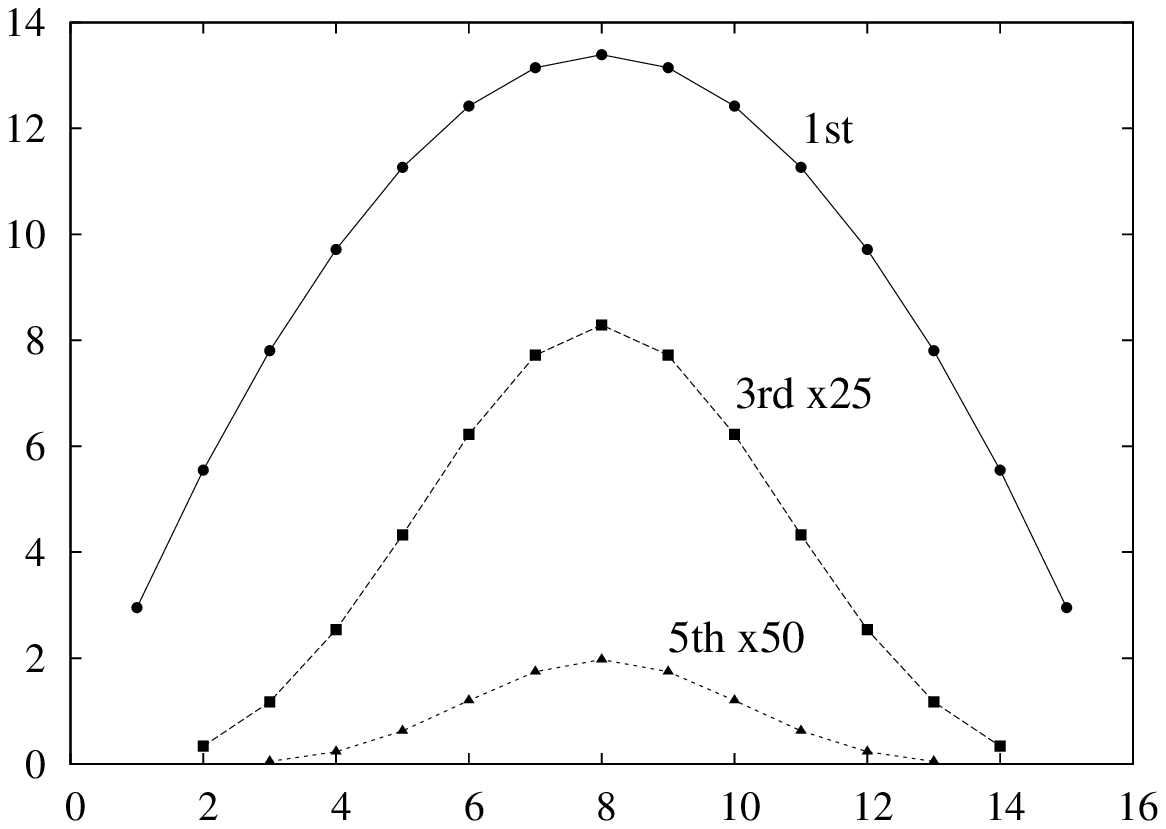}
\includegraphics[scale=.63]{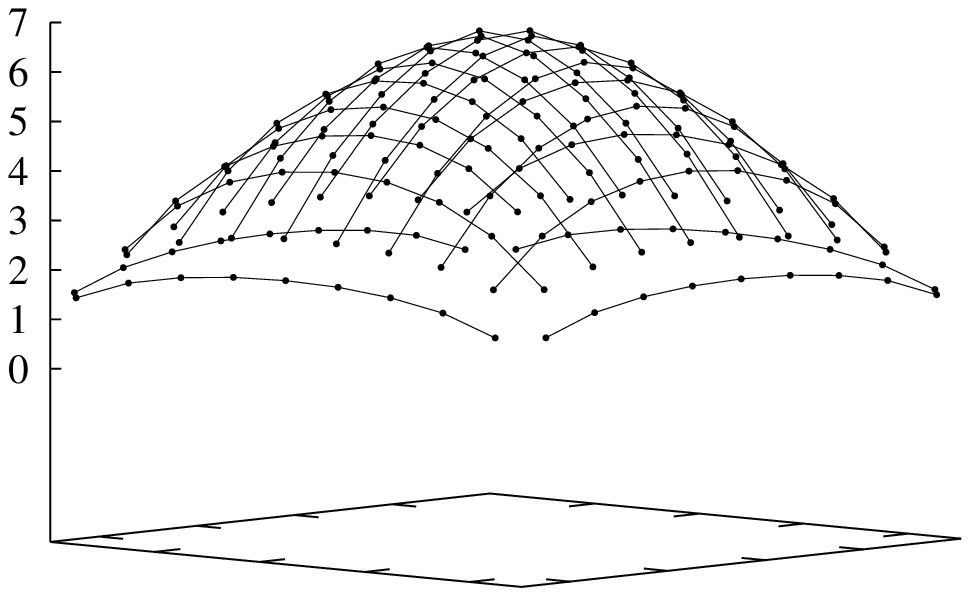}
\caption{Matrix elements in $\mathcal{H}_{\alpha}$ for a hopping model.
 Left: First, third and fifth neighbour hopping in a segment of $L=16$ sites.
Right: First-neighbour hopping in a $10 \times 10$ square.} 
\label{fig:heffij}
\end{figure}

For any finite subsystem, $\mathcal{H}_{\alpha}$ can in principle be determined numerically.
This is particularly simple for the homogeneous hopping model. Then the matrix elements
$h_{i,j}$ follow from the correlation-matrix eigenfunctions via
\begin{equation}
h_{i,j}= \sum_l \phi_l(i) \varepsilon_l \phi_l(j)
\label{matrixh}
\end{equation}
The result for a segment in a chain is shown on the left of Fig. \ref{fig:heffij}.
%%%%%%%%%%%%%%%%%%%%%%%%%%%%%%%%%%%%%%%%%%%%%%%%
%%%%%%%%%%%%%%%%%%%%%%%%%%%%%%%%%%%%%%%%%%%%%%%%
The dominant elements are
those for nearest-neighbour hopping and vary roughly parabolically. This is the 
generalization of the linear law in the semi-infinite chain to this geometry. However, 
there is
also hopping to more distant neighbours, although with rapidly decreasing amplitude.
If the segment is located at the end of a chain, one finds the same behaviour but with 
only one-half of the parabola, i.e. the hopping saturates at the free end. The
situation for a square in a two-dimensional lattice is shown on the right of the
figure. Going parallel to an edge, the hopping in this direction varies again 
parabolically. It is smallest close to the edge and largest halfway in between the
edges. This shows that the inhomogeneity in $\mathcal{H}_{\alpha}$ always follows the same
pattern. One finds it also in the XXZ model with $\Delta=1/2$ \cite{Nienhuis09}.
\par
In the XX chain, one can actually show that $\mathcal{H}_{\alpha}$ for a segment commutes 
with the operator,
\begin{equation}
{\mathcal{T}}= \sum_{i=1}^{L-1} \frac {i(L-i)}{L} \, [c_i^{\dagger} c_{i+1}+ c_{i+1}^{\dagger} c_i]
\label{commutop}
\end{equation}
where the hopping is strictly only to the nearest neighbours and has \emph{exactly}
parabolic form \cite{Peschel04}. Thus they have common eigenfunctions, and the result 
(\ref{continphi}) was actually found from $\mathcal{T}$. Also the low-lying eigenvalues are 
related, and it could be that in the limit $L \rightarrow \infty$ both operators become 
identical up to a factor. One cannot check that numerically, however, because then
large $\varepsilon_l$ appear which are not accessible, see section 3.

\subsection{Definition of a temperature}

It has been pointed out in section 3 that $\rho_{\alpha}$ is not a true Boltzmann
operator, since $\mathcal{H}_{\alpha}$ differs from the Hamiltonian $H$, as shown above.
However, if the single-particle excitations have the same functional form, one can
bypass this argument. This is the case for the homogeneous hopping model
\cite{Eisler/Legeza/Racz06}. Then the $\varepsilon_l$ vary linearly for large $L$ 
according to (\ref{epsilonconf}) and the same holds for
the single-particle energies in $H$ in the vicinity of the Fermi point. For hopping
to nearest neighbours with matrix element $t/2$ these are, in the subsystem, given by 
\begin{equation}
\omega_l= t \frac {\pi}{2(L+1)} (2l-1)
\label{omegalin}
\end{equation}
Therefore, one can write $\varepsilon_l = \beta \omega_l$ with an effective
temperature
\begin{equation}
T= t \pi \frac {\ln L}{L} 
\label{temperature}
\end{equation}
which depends on the length of the subsystem and vanishes for  $L \rightarrow \infty$.
Therefore $\rho_{\alpha}$ can be regarded as a true grand canonical Boltzmann distribution 
for all expectation values, where only the small single-particle energies are important and
the wave functions do not play a role. This holds, for example, for the particle-number
fluctuations in the subsystem, which vary as $T\,
L$ at finite temperatures. Inserting
(\ref{temperature}), this is turned into the $\ln L$-behaviour for the segment in the 
chain.

\subsection{Thermal states}

\begin{figure}[htb]
\centering
\includegraphics[scale=.65]{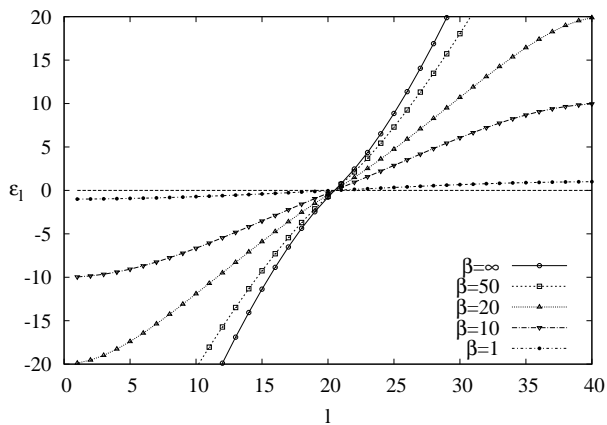}
\caption{Single-particle spectrum as a function of the inverse temperature for a
segment of $L=40$ sites in an infinite hopping model.} 
\label{fig:epstherm}
\end{figure}

Although our interest is in ground-state properties, it is instructive to see what happens
if one calculates $\rho_{\alpha}$ for a system at a finite temperature. This is quite
easy with the correlation function approach, and the resulting spectra for the homogeneous 
half-filled hopping model with $t=1$ are shown in Fig. \ref{fig:epstherm}.
%%%%%%%%%%%%%%%%%%%%%%%%%%%%%%%%%%%%%%%%%%%%%%%%
%%%%%%%%%%%%%%%%%%%%%%%%%%%%%%%%%%%%%%%%%%%%%%%%
The steepest curve is the 
ground-state result. As the temperature is increased it flattens, bends over and assumes
the shape of the dispersion $\omega_q=-\cos q$ for the single-particle energies in $H$.
In fact, one can write, expanding the Fermi function for $\beta = 1/k_B T \ll 1$
\begin{eqnarray}
C_{m,n}= \int_{-\pi}^{\pi} \frac {\mathrm{d}q}{2\pi} \, e^{-iq(m-n)} f(\omega_q) 
\simeq \int_{-\pi}^{\pi} \frac {\mathrm{d}q}{2\pi} \, e^{-iq(m-n)} 
\frac {1}{2}(1+\beta \frac {\cos q}{2}) \nonumber \\
=  \frac {1}{2} \left[\delta_{m,n}+\frac {\beta }{4}(\delta_{m,n+1}+ \delta_{m,n-1}) \right]
\label{corrXXthermal}
\end{eqnarray}
which has eigenvalues in the subsystem
\begin{equation}
\zeta_{l}= \frac {1}{2} (1 + \beta \frac {\cos q_l}{2}) ,\quad q_l =\frac {\pi}{L+1}\,l ,
\quad l= 1,2,...L 
\label{zetathermal}
\end{equation}
and gives 
\begin{equation}
\varepsilon_l = - \beta \, \cos q_l 
\label{epsthermal}
\end{equation}
In other words, for high temperature
\begin{equation}
{\mathcal{H}_{\alpha}} \rightarrow \beta H_{\alpha} 
\label{Heffthermal}
\end{equation}
which is a very plausible result. Apart from the shape of the spectrum, the essential
point is that the level spacing is reduced from a value of order one to $\sim 1/L$.
Such a situation is also found in quenches, see section 7.

\section{Entanglement entropy}

In this section, we show how the properties of the RDM spectra seen in section 4  
translate into in the behaviour of the entanglement entropy.  Due to the form of 
the $\rho_{\alpha}$ it is given by the same expression as in statistical physics
\begin{equation}
S = \pm \sum_l \ln (1 \pm\mathrm{e}^{-\varepsilon_l})+\sum_l
\frac{\varepsilon_l}{\mathrm{e}^{\varepsilon_l} \pm 1}
\label{ent-thermo}
\end{equation}
where the upper(lower) sign refers to fermions(bosons). From this formula, one
can immediately draw two general conclusions
\begin{itemize}
\item 
The largest contributions come from small ${\varepsilon_l}$ (corresponding to
high temperature in usual thermodynamics). Therefore the entropy will be
particularly large in critical systems. For fermions its maximum value of $L \ln 2$ 
is reached if all $\varepsilon_l$ vanish.
\item
If all ${\varepsilon_l}$ are $m$-fold degenerate, the value of $S$ is $m$ times
its value without degeneracy. This answers e.g. how $S$ compares for one or two
(noncritical) interfaces, or for one or two Fermi seas.
\end{itemize}

\par

\noindent (I) One dimension
\par
Analytical results can be given for the non-critical half-chains with the spectrum 
(\ref{epsilonCTM}). The sums then lead to elliptic integrals \cite{Peschel05}
and one obtains for fermions in the disordered region
\begin{equation}
   S=  \frac {1} {24} \left[ \;\ln \left( \frac {16} {k^2 k'^2} \right) + (k^2-k'^2)
         \frac {4 I(k) I(k')} {\pi} \right] ,
   \label{eqn:Sferm1}
\end{equation}
while for bosons the formula is 
\begin{equation}
   S=  -\frac {1} {24} \left[ \;\ln \left( \frac {16 k'^4} {k^2} \right) - (1+k^2)
         \frac {4 I(k) I(k')} {\pi} \right] .
   \label{eqn:Sbos}
\end{equation}
A similar expression with an additional contribution of $\ln 2$ coming from the 
eigenvalue $\varepsilon_0 = 0$ holds in the ordered region. Also the results for
XXZ and XYZ chains \cite{CC04,Ercolessi09} can be brought in this form.
The entropy for the
anisotropic XY chain with $h=0$ can be written as the sum of the expressions
in the ordered and the disordered region \cite{XYPeschel04, Igloi/Juhasz08}.
A plot of $S$, based on a numerical evaluation of the sums, was first shown in
\cite{CC04}. Curves for the XY model can be found in \cite{Franchini07}.
The case of a segment in an XY chain was treated even before the
half-chain. Using the correlation matrix and solving a Riemann-Hilbert
problem, $S$ was obtained as an integral over theta functions \cite{Its05,Its/Korepin09}.
This is equivalent to the half-chain result multiplied by two.
\par
In the disordered region there is little 
difference between fermions and bosons. The values of $S$ are typically of order one or 
smaller, so that the corresponding ground states have $M_{\mathrm{eff}} \sim 1-10$
states in the Schmidt decomposition. This reflects the rapid decay of the spectrum in 
Fig. \ref{fig:spectra_Ising}. 
An exception is only the vicinity of the critical point. As anticipated, $S$ becomes large 
there and actually diverges for this geometry. The formulae give for $k \rightarrow 1$ 
\begin{equation}
   S \simeq \frac {c} {6} \;\ln \left( \frac {1} {1-k} \right)
 \label{eqn:Sdiv}
\end{equation}
where $c=1/2$ for the TI model and $c=1$ for the bosons. Since the correlation length 
varies as $\xi \sim 1/(1-k)$, the logarithm is of the form $\ln(\xi/a)$ \cite{CC04}.
\par
%%%%%%%%%%%%%%%%%%%%%%%%%%%%%%%%%%%%%%%%%%%%%%%%
\begin{figure}[htb]
\centering
\includegraphics[scale=.35,angle=270]{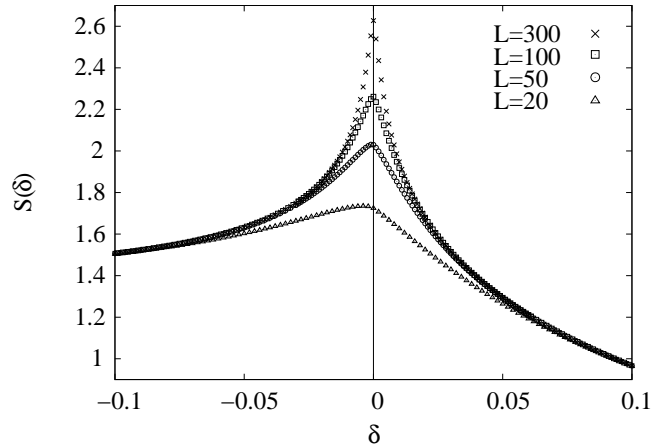}
\caption{Entanglement entropy for segments of different size $L$ in
a one-dimensional hopping model as a function of the dimerization parameter $\delta$.
The development of a singularity in case of vanishing dimerization is clearly visible.
After Ref. \cite{Greifswald08}.} 
\label{fig:entdim}
\end{figure}
%%%%%%%%%%%%%%%%%%%%%%%%%%%%%%%%%%%%%%%%%%%%%%%%
%
The behaviour for a finite subsystem is shown in Fig. \ref{fig:entdim} for segments 
in a dimerized hopping model. In this case, $S$ no longer diverges at criticality but 
shows a maximum which becomes higher with increasing $L$. The size dependence at the
critical point can be obtained in a very simple way \cite{Callan/Wilczek94}. Using
the asymptotic form (\ref{epsilonconf}) of the $\varepsilon_l$ in (\ref{ent-thermo}) and 
converting the sums into integrals gives
\begin{eqnarray}
  S =  \frac {2 \,\ln L}{\pi^2} \, \left [ \, \int_{0}^{\infty} \mathrm{d}\varepsilon \; 
       \ln(1+ \exp(-\varepsilon)) +
       \int_{0}^{\infty} \mathrm{d}\varepsilon \;\frac {\varepsilon}{\exp(\varepsilon)+1} \right ]
\label{eqn:Scrit1}
\end{eqnarray}
and since both integrals equal $\pi^2/12$ one finds  
\begin{equation}
 S = \frac {1}{3} \, \ln L
\label{eq:Scrit2}
\end{equation}
On the lattice, this behaviour was first found numerically \cite{Vidal03,Latorre04}
and then by using the asymptotic properties of the correlation matrices \cite{Jin/Korepin04,
Keating/Mezzadri04}. The general formula for critical chains is
\begin{equation}
 S = \nu \frac {c}{6} \, \ln L + k
\label{eq:Scrit3}
\end{equation}
Here $c$ is the central charge, $\nu$ the number of contact points between the
(singly-connected) subsystem and the remainder of the chain, and $k$ a non-universal
constant which depends on the model parameters and the geometry. An estimate for $k$
can be obtained if one replaces $\ln L \rightarrow \ln L+2.5$ in (\ref{eqn:Scrit1}),
using the scaling found for the first few eigenvalues. This gives $k \sim 0.8$ for
the hopping model, whereas the correct value is $k=0.726$. As the numerics show,
the logarithmic behaviour of $S$ can already be observed in relatively small systems, 
where (\ref{epsilonconf}) is not yet valid. Since it holds for all conformally invariant 
models \cite{Holzhey94,CC04} the formula (\ref{eq:Scrit3}) is a central result.
\par
\begin{figure}[htb]
\centering
\includegraphics[scale=.3,angle=270]{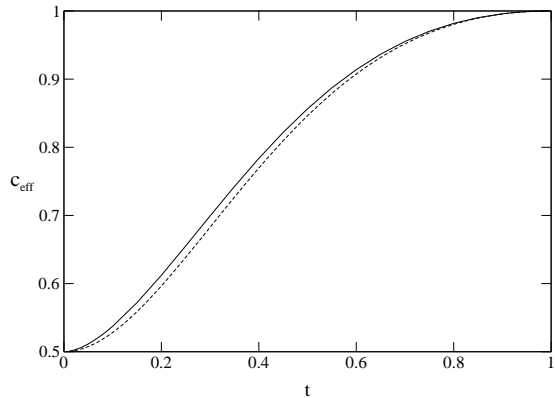}
\caption{Effective central charge for one interface defect in a hopping model as a
function of the defect strength. The dotted curve is an analytical approximation.
After Ref. \cite{Peschel05}.}
\label{fig:ceffdef}
\end{figure}

The interpolation between one and two contact points via a modified bond has already been
discussed in section 4.1. Regarding the entropy, it can be described by an effective 
central charge $c_{\mathrm{eff}}=\nu \, c / 2$ in (\ref{eq:Scrit3}) which varies continuously
between 1/2 and 1. The spectrum on the left of Fig. \ref{fig:eps_defect} then leads to
the result in Fig. \ref{fig:ceffdef}.
%
%%%%%%%%%%%%%%%%%%%%%%%%%%%%%%%%%%%%%%%%%%%%%%%%
%%%%%%%%%%%%%%%%%%%%%%%%%%%%%%%%%%%%%%%%%%%%%%%%
%
A formula for $c_{\mathrm{eff}}$ based on boundary conformal theory was given in
\cite{Sakai/Satoh08}.
The problem was also generalized to the case of two coupled planes \cite{Levine/Miller08}.
Completely inhomogeneous systems were studied in the form of chains with extended defects 
\cite{Igloi/Szatmari/Lin09}, gradients \cite{Eisler/Igloi/Peschel09},
aperiodic \cite{Igloi/Juhasz/Zimboras07} and random \cite{Laflo05,Igloietal07,Igloi/Lin08}
couplings.
On the other hand, one can consider situations where the subsystem is not singly connected 
and thus has many contact points. For comb-like geometries, the leading term in $S$ then 
becomes proportional to $L$ \cite{Keating/Mezzadri/Novaes06}. For example, if the sites of 
the subsystem are two lattice spacings apart, one has $S= L\ln2$ in the hopping model. This
is a direct consequence of (\ref{corrXX}) which reduces to $C_{i,j}=\delta_{i,j}/2$ and
gives $\varepsilon_l=0$ for all $l$. Conformal results for multiple intervals are reviewed
in \cite{CCreview}.
\bigskip\par
\noindent (II) Two dimensions
\par
The influence of the interface on the spectra in two dimensions has already been 
demonstrated in section 4.2. In the entanglement entropy, it leads to the famous 
``area law'' which has been the topic of many investigations, see 
\cite{Eisert/Cramer/Plenio09} for a recent review. Consider, for example, the 
non-critical half-strip of oscillators. Each band of $\varepsilon_{l,q}$ contributes
an amount of order $M$ to $S$ which thereby becomes proportional to the length of
the interface. Expressed differently, $S$ is the sum of the $M$ individual
$q$-chain entropies and can be written, for large $M$, 
\begin{equation}
  S=  \sum_{l,q} s_{l,q} \simeq  M \int_0^{\pi} \frac {\mathrm{d}q}{\pi} \sum_{l} s_{l,q} 
   \label{eqn:Sband}
\end{equation}
For a square-shaped subsystem where the lowest band contains as many states as there 
are interface sites, one obtains an analogous result. 
\par
The argument also holds for critical systems \cite{Cramer/Eisert/Plenio07}. Regarding 
a two-dimensional hopping model as a system of coupled chains, the Hamiltonian reads, 
for $t=1$,
\begin{equation} 
H=- \sum_q  \sum_n \left[ \frac {1}{2}( c_{n,q}^{\dagger} c_{n+1,q}+ 
c_{n+1,q}^{\dagger} c_{n,q}) 
+\cos q \, c_{n,q}^{\dagger} c_{n,q} \right]
\label{coupledhop}
\end{equation}
Thus for each $q$-value one has a chain with a chemical potential $\mu=\cos q$. This
affects the filling but does not change the $\ln L$-behaviour of $S$ , which therefore
becomes proportional to $M \ln L$. This is still an area law, but the occurrence of the
second length disturbs the picture somewhat. The same holds for a square $L \times L$ 
subsystem with the spectrum found in Fig. \ref{fig:eps_xx2d}. There the number 
of states scales as $L$ but the value of the $\varepsilon_l$ as $1/\ln L$. Thus one finds 
logarithmic corrections to the area law. This was proven exactly by constructing
bounds on $S$ \cite{Wolf06,Gioev/Klich06,Farkas/Zimboras07} and an expression for the
prefactor was given in \cite{Gioev/Klich06}. The problem was also investigated numerically in 
two \cite{BCS06,Li06} and three \cite{Li06} dimensions, and the presence of the
logarithm traced to a finite Fermi surface in the system. For bosonic systems, on the
other hand, no logarithmic corrections were found in the critical limit.
\par
Finally, we want to comment briefly on the largest eigenvalue $w_1$ of the RDM, which
has a close relation to $S$. It plays a role in the so-called single-copy entanglement, 
where one asks which maximally entangled state one can reach from an initial state 
\cite{Eisert/Cramer05}. From (\ref{rhogen}) one sees that 
\begin{equation}
w_1 = \frac{1}{Z}\; e^{-E_0}
\label{wmax}
\end{equation}
where $E_0$ is the smallest eigenvalue of $\mathcal{H}_{\alpha}$. This can be evaluated
for the non-critical half-chains in the same way as $S$. For example, putting 
$S_1=-\ln w_1$, one finds in the bosonic case
\begin{equation}
   S_1=  -\frac {1} {24} \left [\;\ln \left( \frac {16 k'^4} {k^2 } \right) - 
         \pi \frac {I(k')} { I(k)} \right ] .
   \label{eqn:S1boson}
\end{equation}
In the critical limit, this diverges as $S$ and one finds that $S_1 \rightarrow  S/2$.
The same holds for fermions \cite{Eisert/Cramer05,Peschel/Zhao05}. One can show that 
this is a general result for conformally invariant systems 
\cite{Peschel/Zhao05,Orus06,Zhou06}.

\section{Entanglement evolution}

In this last chapter we present results on the entanglement evolution after a change of
the Hamiltonian $H_0 \rightarrow H_1$. This can be treated via correlation functions as 
before and leads to interesting phenomena. The simplest case is a quench, where the 
change is instantaneous and generates a unitary time evolution 
$|\psi(t)\rangle = e^{-iH_1t}|\psi_0\rangle$. If $H_1$ is also a free-particle operator, 
the arguments work as before \cite{CC05} and the RDM has the exponential form (\ref{rhogen})
as in equilibrium but with a time dependent operator
\eq{\mathcal{H}_\alpha(t) = \sum_{l=1}^{L} \varepsilon_l(t) f_l^{\dagger}(t) f_l(t).
\label{eq:Hefftime}}
In the case of particle conservation, the eigenvalues $\varepsilon_l(t)$ 
follow again from the restricted correlation matrix, but now taken at time $t$ 
\eq{C_{i,j}(t)=
\langle \psi_0| \, c_i^{\dagger}(t) \, c_j(t) \, |\psi_0 \rangle \, .
\label{eq:corrt}}
Therefore, one only needs to determine the time evolution of the operators
$c_j(t)$ in the Heisenberg picture. In the following we discuss three different 
situations.

\subsection{Global quench}

In a global quench, the system is modified everywhere in
the same way, a situation which can actually be realized in optical lattices
\cite{BDZ08}. Then the initial state becomes a highly excited state
of $H_1$ with an extensive excess energy.
\par
\begin{figure}[thb]
\center
\includegraphics[scale=0.55]{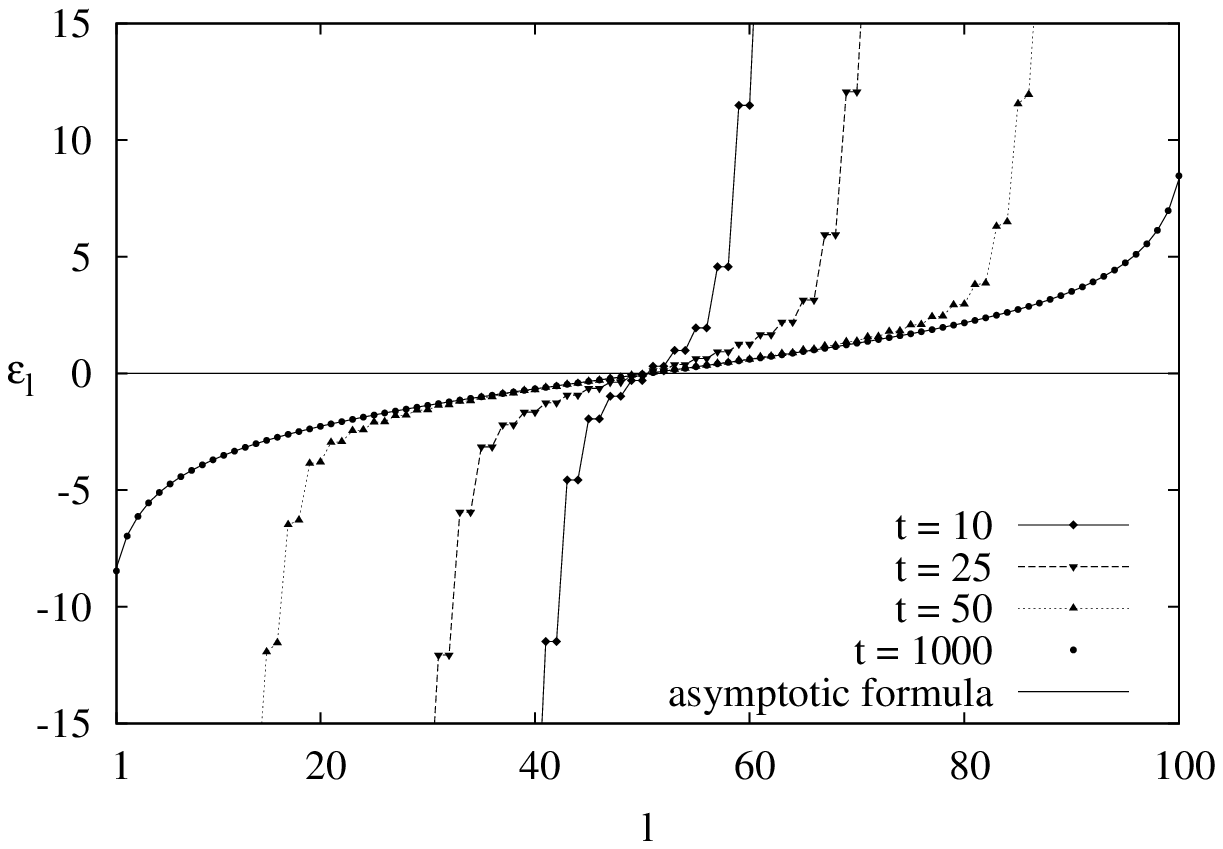}
\includegraphics[scale=0.55]{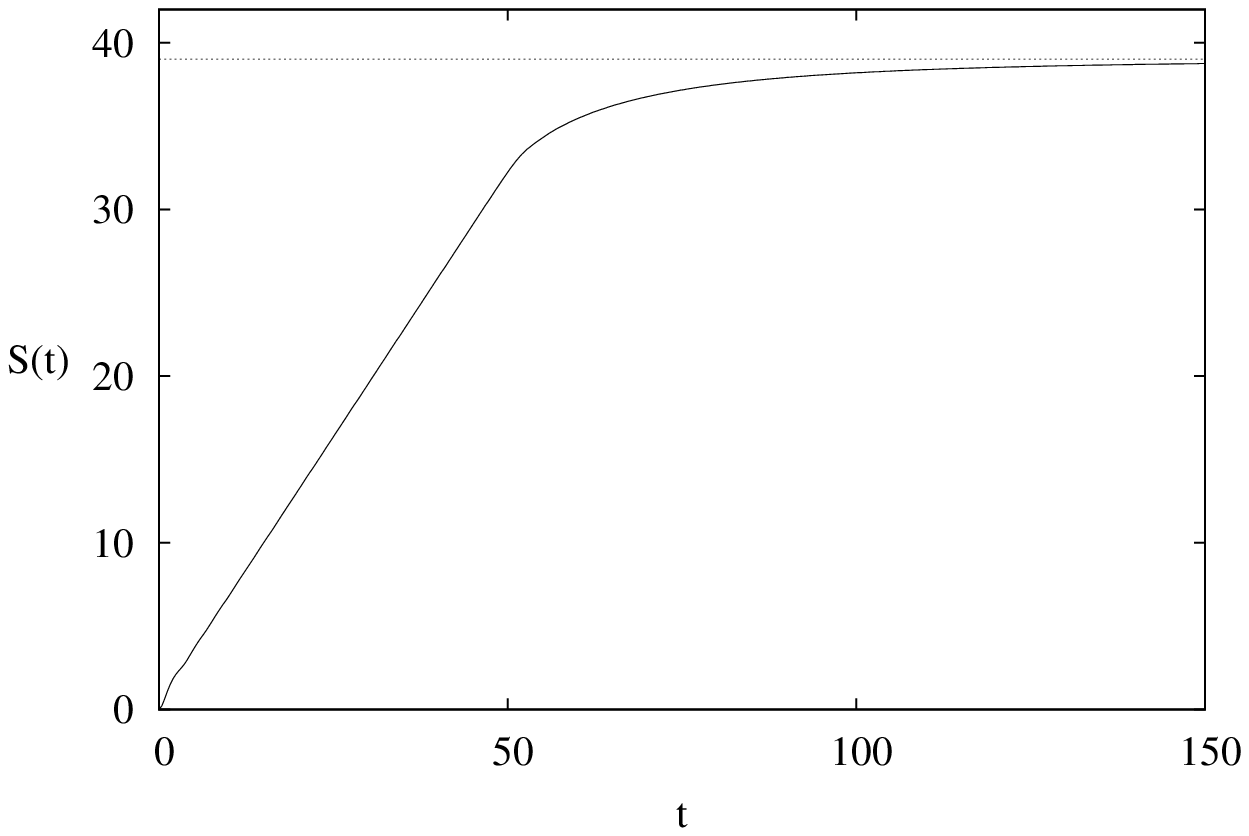}
\caption{Global quench in a hopping model, starting with a fully dimerized initial state.
Left: Time evolution of the single-particle spectrum for a segment of $L=100$ sites. 
After \cite{EP07}. Right: Entanglement entropy with the asymptotic value.}
\label{fig:globalquench}
\end{figure}

An example which illustrates the situation very well is a hopping model
which is initially fully dimerized ($\delta=1$) and then made homogeneous
($\delta=0$). The time evolution of the correlation matrix is then given 
explicitly in terms of Bessel functions \cite{EP07}
\eq{\fl\hspace{1.5cm}
C_{m,n}(t) = \frac {1}{2}  \left[\delta_{m,n} + \frac {1}{2}(\delta_{n,m+1}+
\delta_{n,m-1}) +  e^{-i \frac {\pi}{2}(m+n)} \frac {i(m-n)}{2t}J_{m-n}(2t)
\right] \label{eq:corrt_global}
}
%
%\par
The resulting single-particle spectra are shown on the left of Fig. 
\ref{fig:globalquench}.
One sees that the dispersion is linear near zero and that its slope
decreases with time. This leads to the initial increase of the entropy shown
on the right of the figure. For times $t \gg L/2$, however, the $\varepsilon_l$ 
approach a limiting curve and $S$ saturates. The asymptotic form of the spectrum 
can be obtained from the tridiagonal correlation matrix $C_{m,n}(\infty)$ as in
section 5.4
\eq{
\zeta_l(\infty)= \frac {1}{2} (1+\cos q_l), \;\;\; q_l =\frac {\pi}{L+1}l,
 \;\;\; l=1,2...L
\label{eq:zetainf}
}
leading to
\eq{\varepsilon_l(\infty)= 2 \ln \tan(q_l/2).
\label{eq:epsinf}}
The spacing of the $q_l$ is proportional to $1/L$ and gives an extensive 
entropy $S=L(2\ln 2 - 1)$, a value which was also found in \cite{CC05} for a similar 
quench in the TI model. An initial state where the sites are alternatingly 
full and empty, would even give the maximal possible value $S=L \ln 2$. 
%
%%%%%%%%%%%%%%%%%%%%%%%%%%%%%%%%%%%%%%%
%%%%%%%%%%%%%%%%%%%%%%%%%%%%%%%%%%%%%%%
%
\par
The build-up of an extensive entropy is a typical signature of global quenches. 
It was given a phenomenological description in terms of
emitted pairs of quasiparticles which create entanglement between the subsystem
and the remainder of the system \cite{CC05,CCreview}. In our case these quasiparticles
have maximum velocity $v=1$. This simple picture also
accounts for the ``light-cone effect'' \cite{Bravyi06,Eisert/Osborne06} 
reflected in the entropy at $t \approx L/2$, where the linear increase
turns into a saturation. If one starts from an inhomogeneous state the increase of
$S$ can also be non-linear \cite{Eisler/Igloi/Peschel09}. A closed expression for
$S(t)$ in the XY model was given in \cite{Fagotti/Calabrese08}.
\par
From the extensivity of $S$ one might conjecture a relation of the quench state
to a true thermal state. But a comparison of the spectra in Figs. \ref{fig:epstherm}
and \ref{fig:globalquench} shows that, apart from the linear region, they are different.
A calculation of $\mathcal{H}_{\alpha}(\infty)$ via (\ref{matrixh}) shows that it has
long-range hopping which decreases as $1/|i-j|$ in the interior. However, there 
\emph{are} cases, where the final effective Hamiltonian resembles $H$. 
This happens e.g. if one starts from a chain with alternating site energies $\pm \Delta$. 
Then one finds that for large $\Delta$ the asymptotic $\varepsilon_l$ have the form 
(\ref{epsthermal}) with $\beta=2/\Delta$. This explains the observations in 
\cite{Rigol/Muramatsu/Olshanii06}. In general, the emergence of a $\rho_{\alpha}(\infty)$ 
after a global quench may still be viewed as a local thermalization and is a rather 
general feature of one-dimensional integrable systems, see e.g. 
\cite{Cramer08,Barthel/Schollwoeck08} for a rigorous treatment.

\subsection{Local quench}

A very different behaviour is obtained if one makes sudden local changes in the system,
for example by removing defects in a hopping model. The resulting entanglement evolution
has been investigated for various situations and geometries \cite{EP07,CC07,EKPP08,CCreview}. 
We will consider here the case where a finite segment is joined to an infinite 
half-chain either on one or on both sides \cite{EKPP08}. These two setups will be called 
the semi-infinite and the infinite geometry, respectively.
\par
The time evolution of the Fermi operators $c_n(t)$ is again given in terms of
Bessel functions, and in the infinite geometry the correlation matrix reads
\eq{C_{m,n}(t)=i^{n-m} \sum_{j,l} i^{j-l} J_{m-j}(t) J_{n-l}(t) C_{j,l}(0) \, .
\label{eq:corrt_locinf}}
The double sum over all sites $j,l$ has in this case to be evaluated numerically. 
In the semi-infinite geometry, a similar expression is obtained. 
\par
\begin{figure}[thb]
\center
\includegraphics[scale=0.55]{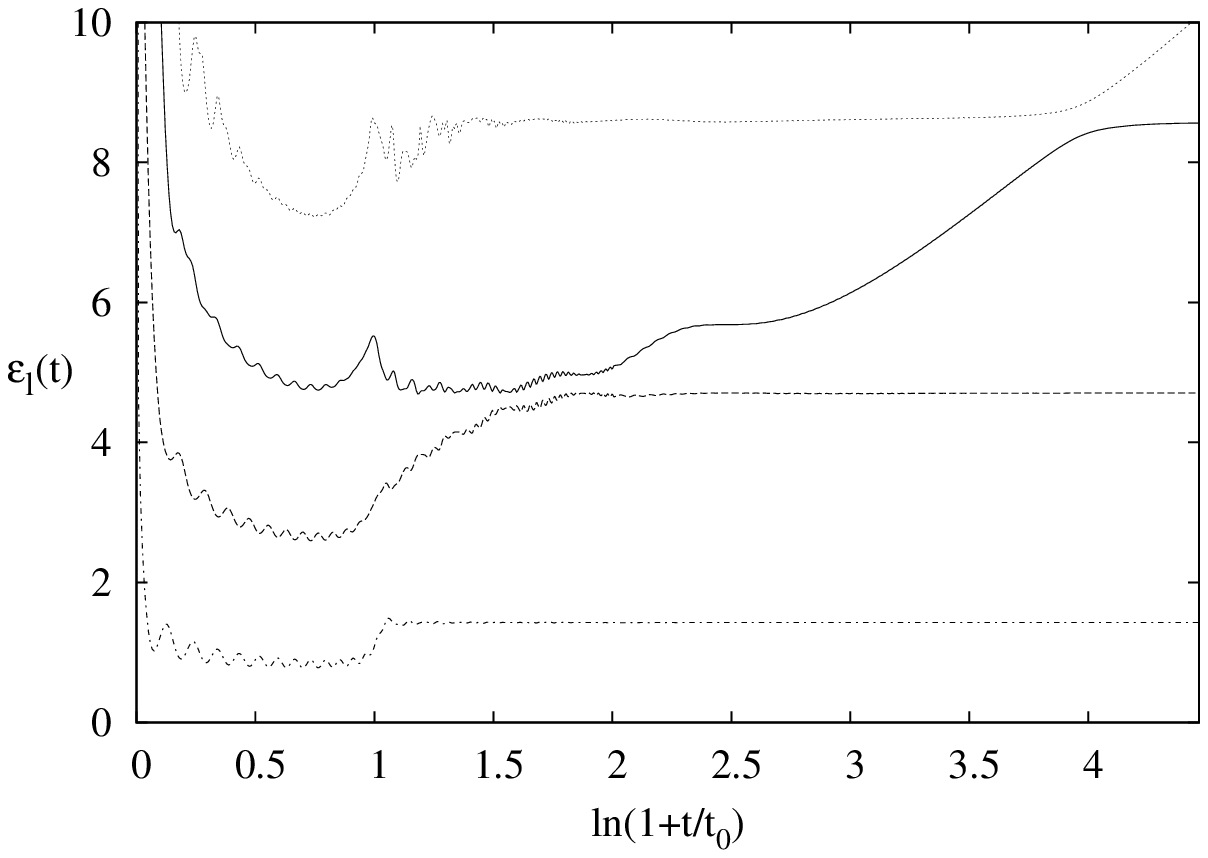}
\includegraphics[scale=0.55]{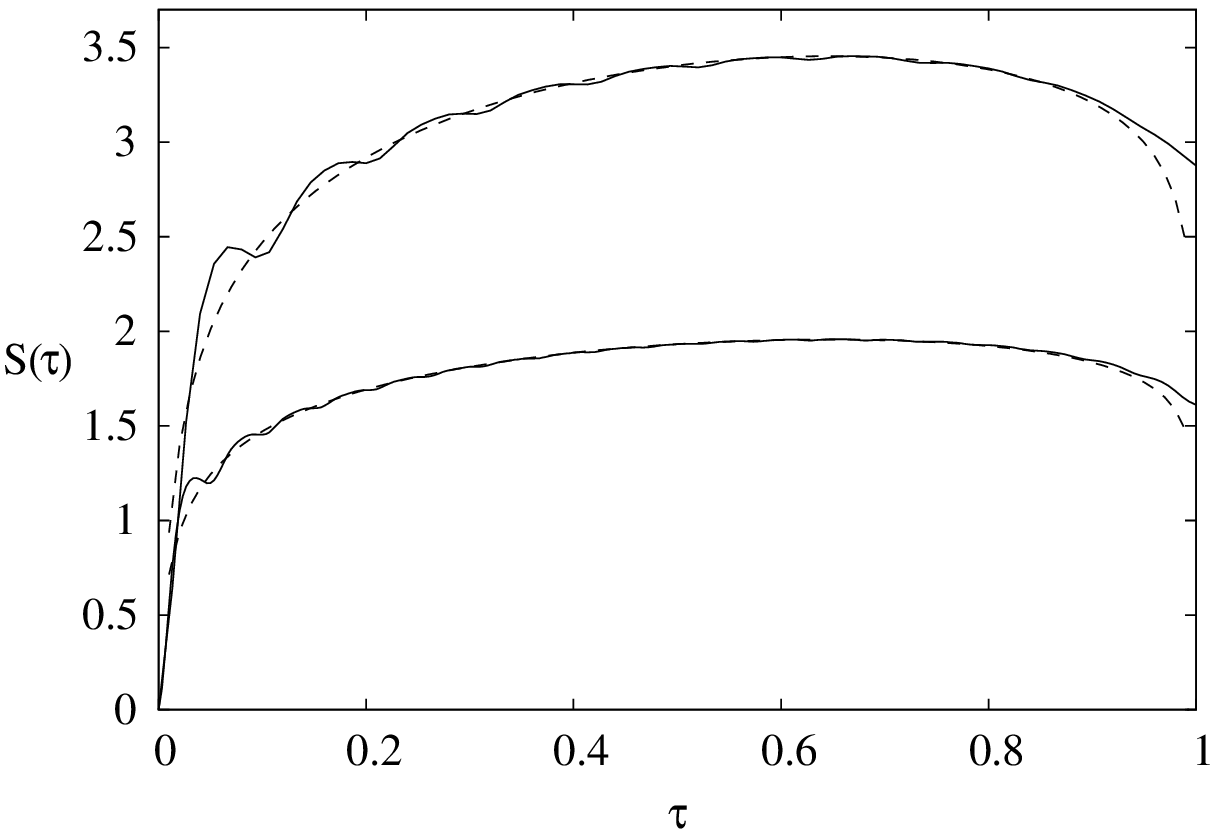}
\caption{Left: Time evolution of the lowest $\varepsilon_l(t)$ for $L=40$. The parameter $t_0$
is chosen such that $t=2L$ gives 1 on the horizontal axis. Right: Entropy evolution in the
rescaled plateau region for $L=60$. Upper curve: Infinite geometry, lower curve: Semi-infinite
geometry. After \cite{EKPP08}.}
\label{fig:epsilon_locsym}
\end{figure}

On the left of Fig. \ref{fig:epsilon_locsym} we show the low-lying single-particle spectrum
for the semi-infinite case on a logarithmic time scale. Since the segment is initially
unentangled, all $\varepsilon_l(0)=\infty$ first drop and evolve to a transient regime up to
$t \approx 2L$ where all but one relax to the stationary eigenvalues of the equilibrium chain. 
The remaining \emph{anomalous} eigenvalue evolves rather slowly showing avoided crossings
with the already relaxed levels. The large-time behaviour is therefore characterized by
a slow approach to the local equilibrium state.
%
%%%%%%%%%%%%%%%%%%%%%%%%%%%%%%%%%%%%%%%%
%%%%%%%%%%%%%%%%%%%%%%%%%%%%%%%%%%%%%%%
%
\par
The resulting entropy evolution in the transient region is shown on the right of 
Fig. \ref{fig:epsilon_locsym}. For both geometries one can see a plateau with a 
characteristic shape but the height and the length are different. The latter effect
is already scaled out in the figure by choosing $\tau=t/L$ for the infinite and $\tau=t/2L$ 
for the semi-infinite case. Using methods of conformal field theory \cite{CC07,EKPP08}, 
one can derive analytical formulae for both cases
\eq{S(t)= \nu \frac c 6 \ln\left[ \frac{4L}{\nu \pi}t
\sin \left(\frac{\nu \pi t}{2L} \right) \right] + k_{\nu}
\label{eq:ent_cft}}
where $\nu$ is the number of contact points and $k_{\nu}$ is a constant which depends 
on the geometry. These curves are indicated by the dashed lines in the figure and, apart 
from deviations at the ends of the interval, are in good agreement with the numerical data. 
For $t \ll L$, Eq. (\ref{eq:ent_cft}) gives a \emph{logarithmic} entropy growth in contrast 
to the \emph{linear} increase in case of the global quench. If $L \rightarrow \infty$ this
behaviour persists for all times.
\par
The emergence of the plateau region can be related to a front starting from the
defect site and propagating with unit velocity. It becomes clearly visible if one looks
at the eigenvectors belonging to the $\varepsilon_l(t)$ in Fig. \ref{fig:epsilon_locsym}.
The plateau ends when the front leaves the subsystem, which also explains the doubling of
the length due to reflection in the semi-infinite case. In addition to these traveling fronts,
which represent the maximal-velocity excitations, there are also more subtle signatures of the
slowest ones. These are visible as flat parts in the evolution of the anomalous eigenvalue. 
\par
In the above examples we have considered defects which initially cut the system into separate
pieces. However, the behaviour is rather similar, if initially the corresponding bonds are 
only weakened. Only the height of the plateau decreases. Since $S(t)$ is proportional to $c$ 
in Eq. (\ref{eq:ent_cft}), this decrease can be described by effective central charges
\cite{EP07}. These depend smoothly on the initial defect strength and one obtains
similar curves as in the equilibrium situation depicted in Fig. \ref{fig:ceffdef}.
A plateau is also found for local quenches in a non-critical TI chain. The difference
in this case is that it becomes flat and does not scale with the subsystem length
\cite{EKPP08}. In summary, for a finite subsystem a local quench is characterized by
bursts of the entanglement: a rapid development of a plateau region is followed by
a slow relaxation towards a local equilibrium.

\subsection{Periodic quench}

As a final example, we discuss a periodic sequence of changes $H_0 \leftrightarrow H_1$ 
and its effect on the entanglement. The change in the Hamiltonian can be either
global or local.
\par
In the global case, we consider again the dimerized hopping model and switch periodically
between dimerizations $\pm \delta$  \cite{EP08,Barmettler08}. This corresponds to a simple 
interchange of weak and strong bonds. The time-evolution operator up to the end of the 
$n$-th period reads
\eq{U(2n\tau)=U^n \quad , \quad
 U = U_0 \, U_1 = \ee^{-i H_0 \tau} \ee^{-i H_1 \tau}
\label{eq:u_perq}}
where $\tau$ is the length of a half-period. For arbitrary times between periods $n$ and $n+1$
one has to multiply $U(2n\tau)$ by an additional unitary operator. Thus, the problem reduces to
finding the diagonal form of $U$ which can be done analytically by a Fourier
transformation. It is convenient to write it as a single exponential of an average
Hamilton operator
\eq{
U= \ee^{-i \bar{H} 2\tau} \quad , \quad
\bar{H} = \sum_q \nu_q (\xi_q^\dag \xi_q - \eta_q^\dag \eta_q)
\label{eq:u_heff}}
with Fermi operators $\xi_q$ and $\eta_q$. In the case $\delta=1$, the single-particle 
energies are given by $\nu_q=\gamma_q/2\tau$ where
\eq{
\cos \gamma_q = \cos^2 \tau - \sin^2 \tau \cos q
\label{eq:gamma_q}}
\par
\begin{figure}[thb]
\center
\includegraphics[scale=0.55]{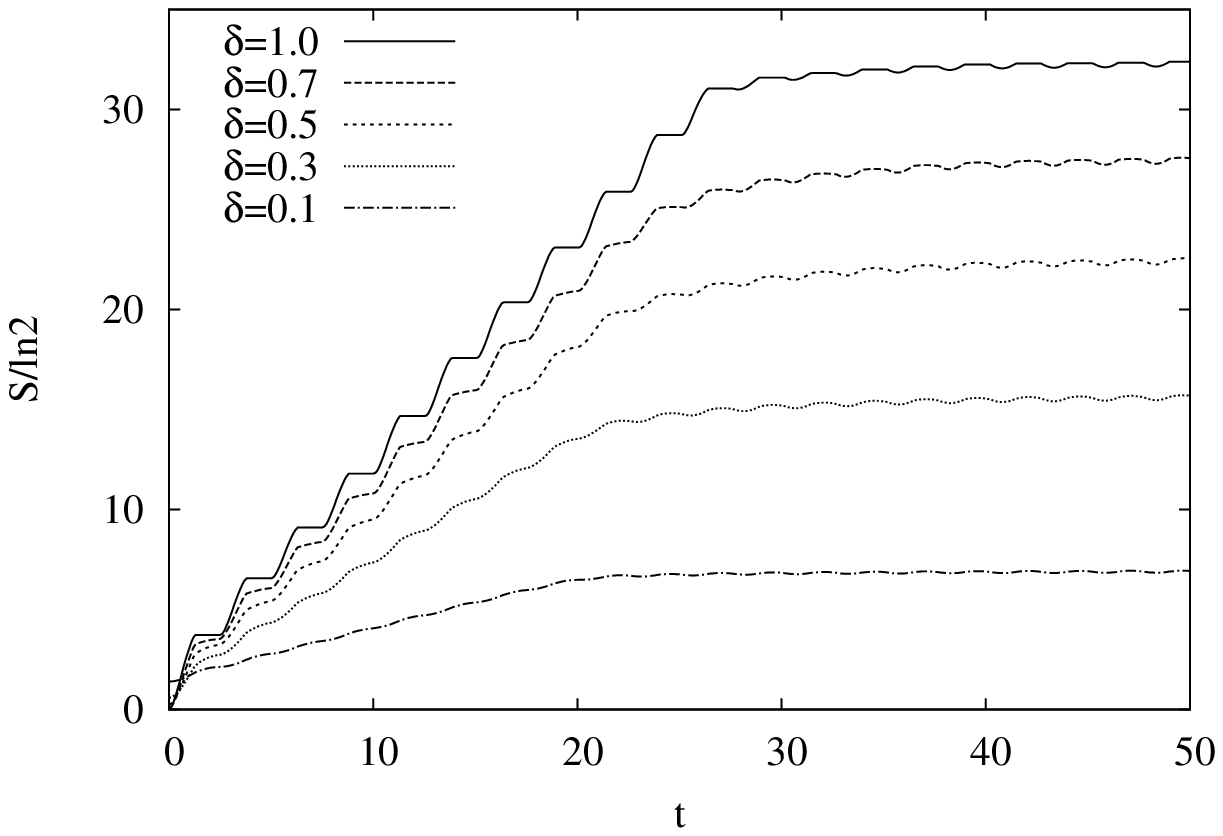}
\includegraphics[scale=0.55]{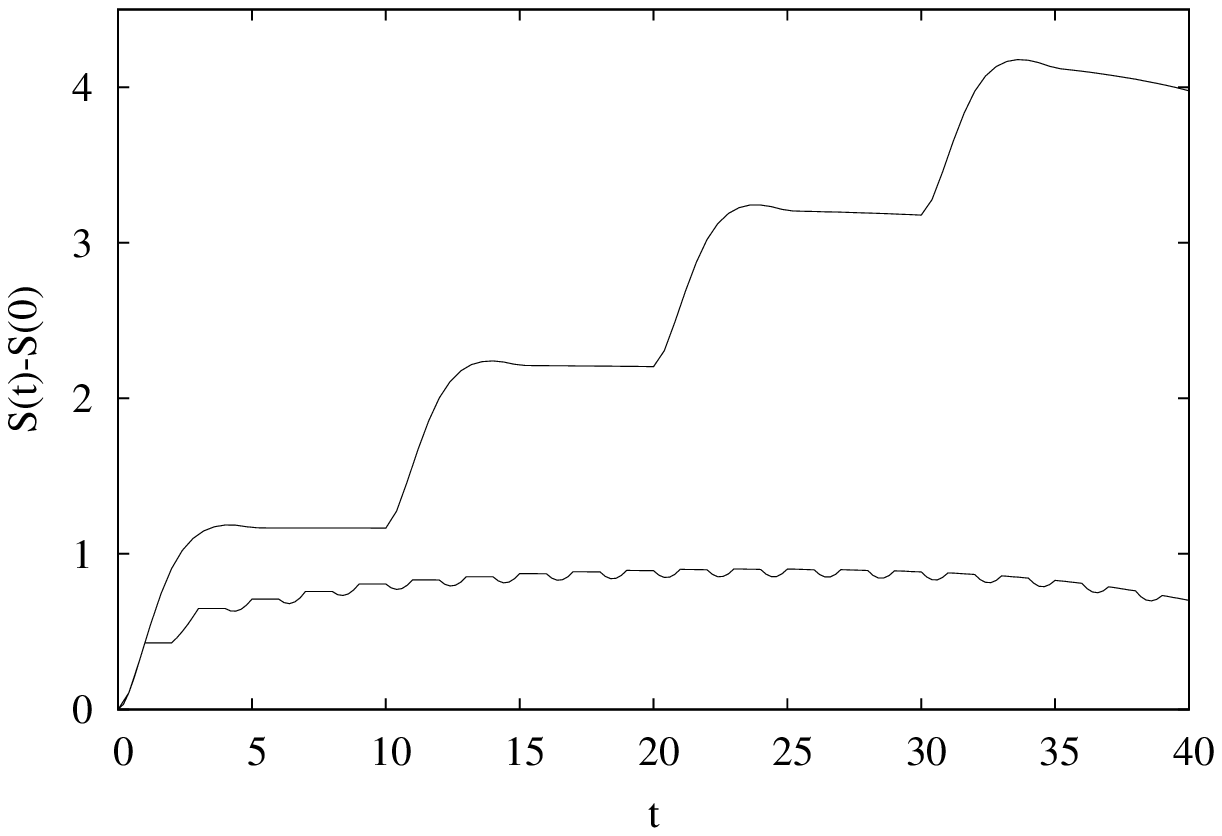}
\caption{Entropy evolution in periodic quenches for $L=40$.
Left: Periodically switched dimerization for $\tau=0.4\pi$ and various $\delta$.
After \cite{EP08}.
Right: Periodically connected chains. Upper curve: $\tau=5$, lower curve: $\tau=1$.}
\label{fig:epsilon_perq}
\end{figure}

The time evolution of the entropy is obtained again from (\ref{eq:corrt}) and depicted on 
the left of Fig. \ref{fig:epsilon_perq} for several values of the dimerization $\delta$ and 
fixed $\tau=0.4 \pi$. The overall behaviour is an initial, step-like increase followed by 
a sharp bend and a final approach to an asymptotic value. The steps are sharp in the fully 
dimerized case, but for smaller $\delta$ they become washed out and their height $\Delta S$ 
decreases. For general $\delta$ and $\tau$ the entropy displays additional slow oscillations.
\par
The characteristics of the time evolution can be understood from the dispersion of
$\nu_q$. For $\delta=1$ and $\tau=\pi/2$ it is strictly linear, resulting in a completely 
regular staircase with $\Delta S = 4 \ln 2$. Thereby a segment of size $L$ becomes maximally 
entangled after $L/4$ periods. This case also gives an \emph{exact} lattice example of the 
quasiparticle picture in \cite{CC05}. In the general case, $\nu_q$ becomes more complicated 
and can have several local maxima, which give rise to the slow oscillations in $S$.
%
%%%%%%%%%%%%%%%%%%%%%%%%%%%%%%%%%%%%%%%%
%%%%%%%%%%%%%%%%%%%%%%%%%%%%%%%%%%%%%%%
%
\par
Apart from the fine structure, the picture is similar to that of the single quench.
Both problems become identical in the limit of very rapid switching, $\tau \to 0$.
Then the average Hamiltonian is just the simple average $\bar{H}=(H_0+H_1)/2 =H$ and 
one recovers the quench to the homogeneous chain. However, the asymptotic entropy
seems to be always larger in the periodic case, and in general is a complicated 
non-monotonic function of $\tau$ \cite{EP08}.
\par
On the right of Fig. \ref{fig:epsilon_perq} we show results for a local periodic
quench. Here two halves of an infinite hopping model are periodically connected and
separated. The subsystem consists of the first $L$ sites in one of the
initially disconnected chains. One sees a characteristic difference. For a large half-period 
$\tau$, one has a step structure like in the case of the global quench, and the entropy 
grows linearly with the number of the periods. This is the result found analytically in 
\cite{Klich/Levitov09} by studying a continuum model and taking the subsystem as one of the
half-chains. For small $\tau$, however, the entropy curve 
resembles the plateau of a single local quench, with an additional fine structure due to the
switching. In this case, the entropy grows only logarithmically. The interpretation is that, 
for slow switching, the system has enough time to recover and thereby the entanglement gain
repeats itself after each new connection. For rapid switching, this is not the case, and for 
$\tau \to 0$ one recovers a single quench as before. The transition between both regimes 
occurs around $\tau=\pi/2$. The phenomenon can also be seen in interacting systems
\cite{Harder08}.

\section{Conclusion}

We have shown that the reduced density matrices of free lattice models have a
special structure. This permits to view entanglement questions in these systems as 
thermodynamic problems and provides a very clear physical picture of the situation. 
In particular, the entanglement entropy can be understood from the character and 
the scaling behaviour of the single-particle spectra, as for conventional 
thermodynamical systems. Therefore the emphasis throughout the review was on 
the properties of these spectra. In addition to presenting them for a 
number of important situations, we also discussed the character of the 
corresponding eigenfunctions and of the effective Hamiltonian itself. Thereby
the role of the interface between the two parts of the system entered in a natural
way. From the character of the eigenfunctions in the ground state problem, one can 
say that the entanglement ``resides'' mainly near the interface \cite{Vidal03}.
Therefore the states 
are rather weakly entangled in one dimension, but already in two dimensions this is no 
longer true and limits the applicability of the DMRG seriously. On the other hand, 
this role of the interface is not a general feature. Not only at finite temperatures,
but also after global quenches, the entanglement entropy becomes extensive and 
typically the whole bulk of the subsystem is involved in the entanglement. On the
other hand, simple local quenches only lead to logarithmic effects and time-dependent 
DMRG can be done. We have only considered quenches, but there are also results
for continuous changes, see e.g. \cite{Cherng/Levitov06,Cincio07}, which one could
discuss in the same way as here. On the whole, time-dependent phenomena should be
the area of further applications. Of course, the study of non-interacting systems is
always combined with the hope that they serve as guides for more realistic ones.
For the DMRG this is certainly the case.

\ack{We would like to thank Pasquale Calabrese for a critical reading of the manuscript.
VE is grateful to the Freie Universit\"at Berlin for hospitality during
his visit where part of this work was carried out. He acknowledges financial
support by the Danish Research Council and QUANTOP.}

\section*{References}

%\bibliographystyle{iopart-num}
%
%\bibliography{review_refs}

\providecommand{\newblock}{}

\end{document}